\documentclass[11pt]{article}

\pdfoutput=1

\usepackage[T1]{fontenc}
\usepackage[latin9]{inputenc}
\usepackage[a4paper]{geometry}
\usepackage[active]{srcltx}
\usepackage{amsmath}
\usepackage{amssymb}
\usepackage{esint}

\usepackage{multicol}
\usepackage{graphicx}
\usepackage{dsfont}
\usepackage{mathtools}
\usepackage{caption}
\allowdisplaybreaks[1]

\usepackage[symbol]{footmisc}

\newcommand{\del}{\partial}
\newcommand{\be}{\begin{equation}}
\newcommand{\ee}{\end{equation}}
\newcommand{\bea}{\begin{eqnarray}}
\newcommand{\eea}{\end{eqnarray}}

\newcommand{\nn}{\nonumber}

\newcommand{\ie}{{\it i.e.}}

\newcommand{\ndt}{\noindent}

\usepackage{tikz}
\newcommand{\xdownarrow}[1]{%
	{\left\downarrow\vbox to #1{}\right.\kern-\nulldelimiterspace}
}

\usetikzlibrary{shapes.geometric, arrows.meta, shadows}

\tikzstyle{arrow} = [thick,->,>=stealth]
\tikzstyle{arrow1} = [thick,<->,>=stealth]
\tikzstyle{line} = [draw, -latex']
\tikzset{
	double arrow/.style args={#1 colored by #2 and #3}{ thick,
		-latex,  line width=1.1*(#1),#2, % first arrow
		postaction={draw,-latex,#3,line width=1.9*(#1/2),
			shorten <=0.7*(#1)/3,shorten >=(#1)/3}, % second arrow
	}
}

\usetikzlibrary{trees}
\usepackage{varwidth}
\usepackage{pifont}
\usepackage{cancel}
\usetikzlibrary{shapes.geometric, arrows, shadows}

\tikzstyle{forces} = [rectangle, rounded corners, minimum width=2cm, minimum height=0.8cm,text centered, draw= gray]
\tikzstyle{spin} = [rectangle, rounded corners, minimum width=1.3cm, minimum height=0.8cm,text centered, draw=white, fill=white]
\tikzstyle{theory} = [ellipse, minimum width=2.4cm, minimum height=0.8cm,text centered, draw=black, fill=blue!6!]
\tikzstyle{arrow} = [thick,->,>=stealth]
\tikzstyle{arrow1} = [thick,<->,>=stealth]
\tikzstyle{line} = [draw, -latex']
\tikzset{
	double arrow/.style args={#1 colored by #2 and #3}{
		-stealth,line width=#1,#2, % first arrow
		postaction={draw,-stealth,#3,line width=(#1/2),
			shorten <=(#1)/3,shorten >=2*(#1)/3}, % second arrow
	}
}

\usetikzlibrary{external}

\definecolor{ggreen}{rgb}{0.0, 0.5, 0.20}
\definecolor{cgreen}{rgb}{0.9, 0.0, 0.0}
\makeatletter
\usepackage{textcomp}

\pdfoutput=1 % if your are submitting a pdflatex (i.e. if you have
\usepackage{jheppub}% for details on the use of the package, please
\usepackage{etoolbox}% http://ctan.org/pkg/etoolbox
    
    \patchcmd{\maketitle}{\@fpheader}{}{}{}

\usepackage{amsfonts}

\title{\boldmath Carroll theories from Lorentzian light-cone theories}

\author[]{Sucheta Majumdar}

\affiliation{Aix-Marseille Univ, Universit\'e de Toulon, CNRS, \\Centre de Physique Th\'eorique, 13288 Marseille, France}
\emailAdd{sucheta.majumdar@cpt.univ-mrs.fr}

\preprint{}

\abstract{We derive Carrollian field theories via null reduction from Lorentzian light-cone actions in Minkowski spacetime. By suitably deforming the light-cone action, we reduce the Poincar\'e invariance to a Bargmann subgroup, from which both magnetic and electric Carroll actions can be obtained in one lower dimension. Through a canonical analysis, we show that the second-class constraints usually found in Lorentzian light-cone theories are absent for these deformed Bargmann-invariant actions.  We demonstrate the procedure for theories with and without gauge symmetry. Notably, while the magnetic Carroll sector can be directly derived from the original Lorentzian action, the deformation is essential to obtain the electric Carroll sector. We further argue that magnetic Carroll solutions in $d$ dimensions represent a consistent truncation of the solutions of the $(d+1)$-dimensional Lorentzian parent theory, providing an effective description of light-cone dynamics near a null hypersurface. For gauge theories, we also highlight the role of the light-cone gauge condition in deriving Carrollian theories.} 
\makeatother

\begin{document}
\maketitle \flushbottom

\newpage{}

\section{Introduction}
The light-cone formulation provides a powerful framework for studying relativistic field theories by choosing coordinates adapted to null directions~\cite{Dirac:1949cp}. Field theories formulated in light-cone coordinates often reveal interesting structures, with certain modes becoming non-dynamical or constrained, thereby simplifying canonical quantization and the analysis of on-shell physics~\cite{Heinzl:2000ht, Gorsky:2005sf, Ananth:2007zy}. In particular, the light-cone gauge $-$ commonly used in quantum field theories $-$ allows us to eliminate the redundant components of the gauge field, making the distinction between propagating and radiative modes more transparent~\cite{Kogut:1969xa, Brodsky:1997de}. Underlying this formulation is a hidden Galilean invariance in one lower dimension~\cite{Susskind:1967rg,Gomis:1978mv} which has proven extremely useful for tackling many challenging problems in quantum field theory and string  quantization~\cite{Weinberg:1966jm, Brink:1982wv, Green:1982sw, Banks:1996vh}.
\par
Beyond these Galilean or non-relativistic aspects, light-cone field theories also exhibit Carrollian or ultra-relativistic features. This can be attributed to the double-null nature of the light-cone frame resulting in a rich kinematical Lie algebra structure. In a recent work, we revisited the light-cone formulation of quantum field theory and explored various kinematical subalgebras within its Poincar\'e algebra, with particular emphasis on the Carroll, Bargmann, and Galilean subalgebras~\cite{Majumdar:2024rxg}. The Bargmann subalgebras emerge naturally due to the presence of two null directions within the light-cone system, thereby unifying the codimension-one Galilei and Carroll subalgebras within the  light-cone algebra~\cite{Duval:2014uoa, Figueroa-OFarrill:2022pus}.
\par 
Kinematical Lie algebras, such as Galilei, Carroll or Bargmann algebras, describe the algebraic structure of symmetry transformations that govern spacetime kinematics and the evolution of physical systems~\cite{Levy-Leblind:1965, SenGupta:1966qer, Bacry:1968zf, Henneaux:1979vn, Figueroa-OFarrill:2022nui}. The Carroll group $-$ the ultra-relativistic limit of the Poincar\'e group $-$ plays a crucial role in the asymptotic structure of spacetime, black hole horizons, and various other physical settings with applications, range from flat space holography and string theory to condensed matter systems and beyond~\cite{Donnay:2019jiz, Herfray:2021qmp, Adami:2023wbe,Donnay:2022aba, Donnay:2022wvx, Bagchi:2022emh, Bagchi:2023cen, Ciambelli:2018xat, Ciambelli:2018wre, Ciambelli:2019lap, Cardona:2016ytk, deBoer:2023fnj, Oling:2022fft, Blair:2025prd,Blair:2025nno, Bidussi:2023rfs,Bidussi:2021nmp,Perez:2022kax,Figueroa-OFarrill:2023vbj,Bagchi:2022eui}.
\par
There are several ways to formulate field theories in these non-Lorentzian spacetimes. One common approach is the group contraction procedure, where one takes the Carrollian ($c \rightarrow 0$) or Galilean ($c \rightarrow \infty$) limit of a Poincar\'e-invariant action in a given spacetime dimension $d$, with $c$ being the speed of light~\cite{LeBellac:1973unm, Bergshoeff:2017btm, Hansen:2021fxi, Henneaux:2021yzg, Hartong:2022lsy}. Another approach involves formulating Carrollian theories intrinsically from the knowledge of the relevant geometric structures, such as the degenerate metric and null vector field~\cite{Andringa:2010it, Hartong:2015xda, Figueroa-OFarrill:2022mcy} (see also~\cite{Bergshoeff:2022eog} and references therein). A third method, which has been relatively less explored in this context, is the null reduction procedure for deriving Carrollian field theories~\cite{Duval:2014uoa}. In this method, starting with a Bargmann spacetime in $(d+1)$ dimensions, one obtains $d$-dimensional Galilean theories via Kaluza-Klein reduction along a null Killing vector, or Carrollian theories by projecting onto a null hypersurface. The problem of constructing Galilean theories via Kaluza-Klein reduction along a null Killing vector was explored in~\cite{Julia:1994bs, Bagchi:2022twx}.
\par
In an earlier work, Carrollian field theories were constructed using a null reduction approach from Bargmann-invariant actions~\cite{Chen:2023pqf, Sharma:2025rug}. However, it was shown there that the electric and magnetic sectors of Carrollian theories originate from \textit{different} Bargmann-invariant actions. This implies that, in general, one needs to construct two separate Bargmann actions in the higher-dimensional spacetime $-$ one for the electric sector and another for the magnetic sector. Furthermore, this stands in contrast with the group contraction method, where both electric and magnetic Carrollian actions follow from the same Lorentzian action but arise from different rescalings of the fields as $c \rightarrow 0$.
\par
The goal of this paper is to derive $d$-dimensional Carroll-invariant actions using the null reduction method, starting from a \textit{single} Lorentzian light-cone action in $(d+1)$ dimensions. We discuss several illustrative cases, including point particle actions, scalar fields and electromagnetism to highlight the features of gauge systems, Yang-Mills theory to demonstrate interacting cases, and finally the free $p$-form fields to generalize our method. Since in this work we focus only on light-cone theories in Minkowski spacetime, we are essentially confining ourselves to the flat Bargmann case.  We present the derivation of Carrollian theories in both Hamiltonian and Lagrangian formulations. As light-cone theories are first order in time and involve second-class primary constraints, it is instructive to perform a rigorous canonical analysis to better understand the subtleties involving the constraint structure and gauge fixing.
\par 
The outline of the paper is as follows. We begin with a brief summary of the key features of light-cone Minkowski spacetime and its corresponding Poincar\'e algebra. In the next section, we outline the null reduction procedure using the example of point particles in the Hamiltonian framework. We then demonstrate the need to deform the Lorentzian light-cone action into a more general Bargmann-invariant action, which yields both the electric and magnetic sectors upon null reduction. In Section 3, we proceed to derive Carroll field theories via the null reduction method, starting with free scalars. Section 4 is devoted to the discussion of gauge theories. Starting with electromagnetism, we focus on the constraint analysis of light-cone gauge theories and then detail the procedure to derive electric and magnetic Carroll actions from the light-cone Maxwell action. We also discuss how the gauge-fixed formulation differs from the standard case, where the gauge freedom is left intact or manifest. Having established the general method, we then show how to obtain the same Carrollian actions directly in the Lagrangian formulation, without going through the detailed Dirac-Bergmann analysis of constraints~\cite{Dirac:1950pj,Bergmann:1949zz}. This allows us to easily extend the procedure to more involved and interesting cases such as Yang-Mills theory and $p$-form fields. We conclude with remarks on an intriguing connection between the light-cone formulation and magnetic Carroll theories, along with some future directions.

\section{Bargmann structure of light-cone Minkowski }
In this section, we summarize the key features of the light-cone frame and its corresponding Poincar\'e algebra, that are relevant to Carrollian physics.
\par
In $(d+1)$ dimensions, the light-cone coordinates $x^\mu = (x^+,x^-,x^i)$ are defined as
\begin{eqnarray} \label{LCcoord}
x^+ = \frac{1}{\sqrt 2}(x^0 + x^d) \,,&& x^-= \frac{1}{\sqrt 2}(x^0 - x^d) \,, \quad i,j = 1,2, ..., d-1\,.
\end{eqnarray}
The line element in Minkowski spacetime reads
\begin{equation} \label{LCds2}
dS^2= \eta^{lc}_{\mu\nu} dx^\mu dx^\nu = -2 dx^+ dx^- + \delta_{ij}dx^i dx^j\,.
\end{equation}
It is clear from \eqref{LCds2} that Minkowski spacetime in light-cone coordinates has a flat Bargmann structure. There exist two vector $\mathfrak n =\del_+$ and $\mathfrak m= \del_-$, which satisfy the property
\be
\mathfrak n^\mu \mathfrak n^\nu \eta^{lc}_{\mu\nu} = \mathfrak m^\mu \mathfrak m^\nu \eta^{lc}_{\mu\nu} = 0 \,.
\ee 
As a result, among the generators of the light-cone Poincar\'e algebra $\mathfrak p$, that we relabel as 
\begin{eqnarray} \label{Poincare}
P_+ = E\,,&  P_- = \eta\,, & ~~P_i \,, \nn\\
M_{+i} = K_i\,, & ~~M_{-i} = B_{i}\,,&~~ M_{ij}\,, \quad M_{+-}= D \,.
\end{eqnarray}
one can find the two subsets that form two distinct Bargmann subalgebras~\footnote{We use the subscript `$+$' or `$-$' to label the algebras, in order to specify which light-cone coordinate, $x^+$ or $x^-$, is chosen as the Newtonian time. In other words, it depends on whether we choose $\mathfrak n$ or $\mathfrak m$ as the preferred null direction to be preserved alongside the light-cone metric. }
 \bea
\mathfrak b_+ &=& \{E, \eta, M_{ij}, B_i, P_i\} \,, \quad \eta \ \text{central element} \,,\\
\mathfrak b_- &=& \{E, \eta, M_{ij}, K_i, P_i\} \,, \quad E \ \text{central element} \,.
\eea
The subalgebra $\mathfrak b_+$ corresponds to the isometries preserving the light-cone metric as well as the null vector $\mathfrak m^\mu$
\be
 L_{\mathfrak b_+} \mathfrak \eta = 0\,, \quad L_{\mathfrak b_+} \mathfrak m =0 \,,
\ee
while the other subalgebra $\mathfrak b_-$ preserves the light-cone metric and the other null vector $\mathfrak n^\mu$
\be
 L_{\mathfrak b_-} \eta = 0\,, \quad L_{\mathfrak b_-} \mathfrak n =0 \,.
\ee
Within each of these Bargmann algebras, we have one Galilei $\mathfrak g_\pm$ and one Carroll $\mathfrak c_\pm$ subgroup in $d$ dimensions. These subalgebras correspond to the isometry group of the Galilean and Carrollian spacetimes that may be obtained from the corresponding Bargmann manifolds via null reduction method~\cite{Duval:2014uoa,Figueroa-OFarrill:2022pus}. In appendix \ref{null-reduce}, we detail the null reduction procedure adapted to light-cone coordinates.
\par 
We summarize these subalgebras in Table~\ref{table:subgroups}, highlighting the central elements and key commutators that define them, while all other commutation relations can be found in~\cite{Majumdar:2024rxg}.  
\begin{table}[h]
 \begin{center}
\caption{Kinematical Lie subalgebras within light-cone Poincar\'e algebra}
\vskip 0.1cm
\label{table:subgroups}
 	\small
	\begin{tabular}{c |c| c } 
		\hline
		Time & $x^+$ Newtonian, $x^-$ Carrollian &$x^+$ Carrollian, $x^-$ Newtonian \\ [0.5ex] 
			\hline\hline
		Bargmann $\mathfrak b$ &$\mathfrak b_+ = \{E, \eta, M_{ij}, B_i, P_i\}$ &  $\mathfrak b_- = \{\eta, E, M_{ij}, K_i, P_i\}$ \\
		& $\eta$ central element & $E$ central element \\
		\hline
		 Galilei $\mathfrak g$ &$\mathfrak g_+ = \{E, M_{ij}, B_i, P_i\} $  & $\mathfrak g_- = \{\eta, M_{ij}, K_i, P_i\} $\\
		 & $[E, B_i]= -P_i$& $[\eta, K_i]= -P_i$\\
		 & Hamiltonian $H_{\mathfrak g_+} = E$& Hamiltonian $H_{\mathfrak g_-} = \eta$\\
		 \hline 
		 Carroll $\mathfrak c $& $\mathfrak c_+ =\{\eta, M_{ij}, B_i, P_i\}$ &   $\mathfrak c_- =\{E, M_{ij}, K_i, P_i\}$  \\
		 &  $[B_i, P_j]= \delta_{ij}\eta$ & $[K_i, P_j]= \delta_{ij}E$ \\
		 & Hamiltonian $H_{\mathfrak c_+} = \eta$ &  Hamiltonian $H_{\mathfrak c_-} = E$ \\
		 \hline 
		\end{tabular}
\normalsize
\end{center}
\end{table}
\par
The two null coordinates, $x^+$ and $x^-$, are fully interchangeable within the light-cone coordinate system. This \textit{choice of the light-cone time} is mirrored in the structure of the subgroups, as exchanging $x^+$ with $x^-$ effectively swaps the roles of certain generators in $\mathfrak p$
\bea
x^+ \longleftrightarrow x^-\,, \quad E \longleftrightarrow \eta\,, \quad K_{i} \longleftrightarrow B_i \,,
\eea
which results in the two sets of Bargmann, Galilei and Carroll subgroups being mapped into each other
\be
\mathfrak b_+ \longleftrightarrow \mathfrak b_-\,, \quad \mathfrak g_+ \longleftrightarrow \mathfrak g_- \,,\quad\mathfrak c_+ \longleftrightarrow\mathfrak c_{-} \,.
\ee
In~\cite{Bagchi:2024epw}, this duality was attributed to the difference between `focusing' on the null directions, which leads to the Carroll subalgebras, versus `focusing' on the transverse directions, which leads to the Galilean subalgebras.
\par
The key point here is that there exist subspaces within light-cone Minkowski spacetime that realize these non-Lorentzian subalgebras as their isometry algebras. Therefore, starting with a Lorentzian theory in the light-cone framework, we can consistently perform a null reduction to one lower dimension, where the spacetime symmetry is given by one of the Carroll subalgebras $\mathfrak c_{\pm}$ within the Poincar\'e algebra $\mathfrak p$. In this paper, we detail the procedure for both worldline actions and field theories.
\section{Point particles}
We first demonstrate our procedure for obtaining Carroll theories through the example of a point-particle action. We present two methods to derive Carroll particle actions via the null reduction approach; however, only one of these methods yields the electric Carroll sector. We highlight the central role of constraints in the Hamiltonian formulation in distinguishing between the magnetic and electric Carroll theories.
\subsection{From Lorentzian light-cone action}
We consider the canonical action of a relativistic point particle in $(d+1)$-dimensional Minkowski spacetime in light-cone coordinates\footnote{Here, the time derivatives are taken with respect to $\tau$, \ie\,, $\dot{x}^\mu (\tau) = \frac{\del x^\mu}{\del \tau}$}
\be \label{PP_can}
\mathcal S [x^\mu, p_\mu, e] = \int d\tau [p_\mu \dot{x}^\mu - \frac{e}{2} (\eta^{\mu \nu}p_\mu p_\nu +m^2) ] \,.
\ee
The arbitrary time parameter $\tau$ is taken to be dimensionless and $e$ represents the einbein field corresponding to the metric of the worldline $\eta_{\tau \tau}$. The Hamiltonian reads
\be \label{PP Ham}
H = \frac{1}{2}\, (\eta^{\mu \nu}p_\mu p_\nu +m^2) \,.
\ee
The momenta $p_\mu$ are canonically conjugate to the coordinates $x^\mu$ with the Poisson bracket relations
\be \label{PP PB}
\{ x^\mu, p_\nu\} = \delta^\mu_{\nu}\,, \quad \{x^\mu, x^\nu\} = 0 \,, \quad \{p_\mu, p_\nu\} = 0\,.
\ee
We can also define a conjugate momenta for the einbein 
\be
\pi_e = \frac{\delta L}{\delta \dot{e}} = 0\,,
\ee
satisfying the Poisson bracket $\{ e, \pi_e\} = 1$. This amounts to a primary constraint, whose stability leads to the on-shell constraint
\be
 \eta^{\mu \nu}p_\mu p_\nu +m^2 = 0\,.
\ee
%Alternatively, the on-shell constraint may also be obtained as the equation of motion with respect to $e$, which acts as a Lagrange multiplier in the canonical action.
\par
In the light-cone coordinates, this action takes the explicitly form
\be \label{PP_can LC}
\mathcal S^{Lor} [x^\mu, p_\mu, e] = \int d\tau\, [p_+ \dot{x}^+ + p_- \dot{x}^- + p_i \dot{x}^i - \frac{e}{2} (-2p_+ p_+ + p_i p^i +m^2) ]\,.
\ee
We refer the reader to~\cite{Zweibach} for more details, particularly for a Lagrangian treatment of the relativistic point particle action in the light-cone gauge. Here, we shall not fix the reparameterization symmetry, as we are interested in the general class of Carroll particle actions that can be derived from the Lorentzian action.
\par
The equations of motion derived from the action above are
\be \label{PP eom}
\dot{x}^\mu - e \,\eta^{\mu \nu} p_\nu = 0 \,, \quad \dot{p}_\mu = 0 \,, 
 \ee
and the mass-shell constraint, which appear as the equation of motion for the Lagrange multiplier $e$
 \be
 \quad p_\mu p^\mu + m^2 = 0\,.
 \ee
 From the first equation in \eqref{PP eom}, we see that the velocity $\dot{x}^+$ is related to the momentum in the $x^-$ direction, $P_-$ and the velocity $\dot{x}^-$ to $P_+$. This point will become important when we perform the null reduction to obtain Carrollian particle actions.
 \par
 We now focus on the spacetime symmetries of the particle action. 
 %Under an infinitesimal light-cone Poincar\'e transformation of the form $\xi^\mu = \omega^\mu_\nu x^\nu + a^\mu$, the coordinates and the momenta transform as
% \be
% \delta x^\mu =  \omega^\mu_\nu x^\nu + a^\mu\,, \quad \delta p^\mu =  \omega^\mu_\nu p^\nu 
% \ee
The generator of infinitesimal Poincar\'e transformations has the form
 \be
 G = \frac{1}{2} \omega^{\mu\nu}M_{\mu \nu} + a^\mu P_{\mu}\,,
 \ee
 where $\omega^\mu_\nu$ and $a^\mu$ are the Lorentz and spacetime translation parameters respectively, and 
 \be
 M_{\mu \nu} = x_{\mu} p_{\nu}-  x_{\nu} p_{\mu}\,, \quad P_\mu = p_\mu\,.
 \ee. 
 The transformation laws then follow from the Poisson brackets
 \bea
&& \delta x^\mu = \{ x^\mu, G\} =  \omega^\mu_\nu x^\nu + a^\mu\, ,\\
 &&  \delta p^\mu = \{ p^\mu, G\} =  \omega^\mu_\nu p^\nu \,.
 \eea
 \par The time reparameterization symmetry, $\tau \to \tau' = \tau + \lambda(\tau)$, is generated by the on-shell constraint
 \be
 G_\lambda = \int d\tau \left[\frac{1}{2}\lambda (\eta^{\mu \nu}p_\mu p_\nu +m^2) + \dot{\lambda} \pi_e\right]\,.
 \ee
 Under the time reparameterization, the einbein and the coordinates transform as
 \bea \label{PP time rep}
&& \delta_\lambda e= \{ e, G_\lambda\}  =\dot{\lambda} \,, \\
&& \delta_\lambda x^\mu= \{ x^\mu, G_\lambda\}  =\lambda p^\mu \,,
 \eea
 while the momenta $p_\mu$ and $\pi_e$ are gauge-invariant.
 \par
 As argued before, the light-cone Minkowski spacetime is a special case of Bargmann spacetimes. Thus, we can perform a null reduction along one of the lightlike directions, say $x^-$, to obtain the action of a point particle in a $d$-dimensional Carroll spacetime. In this context, by null reduction we mean that the worldline of the particle is now restricted to satisfy $x^- (\tau) = x^-_0$, where $x^-_0$ is a constant, or equivalently $\dot{x}^-(\tau) = 0$.
 \par
 By imposing $\dot{x}^- = 0$ in the Lorentzian particle action\eqref{PP_can LC} and then relabelling some of the momentum variables as\footnote{Note that in the light-cone formulation, the energy E is given by $p^- = (p^0 -p^d)/2 = -p_+$}
 \be
 p_+ = - E\,, \quad p_- = \frac{\chi}{e} \,,
 \ee
 we readily obtain a Carrollian action in $d$ dimensions 
 \be
 {}^{(d)} \mathcal S_{Carr} [x^+, x^i,  p_i, E, \chi, e] = \int d\tau [-E \dot{x}^+ + p_i \dot{x}^i - \frac{e}{2} ( p_i p^i + m^2) + \chi E]\,.
 \ee
 The momenta conjugate to $x^-$ is no longer a dynamical variable and instead appears in the action as a Lagrange multiplier $\chi$. The Lagrange multipliers $\chi$ and $e$, impose the two constraints
 \be
 E= 0\,, \quad p_i p^i +m^2 = 0
 \ee 
 which correspond, respectively, to the zero-energy and the on-shell condition for a magnetic Carroll particle. We see that the action so obtained is identical to the canonical action for a Carroll magnetic particle obtained from $c\to 0$ limit of its relativistic counterpart~\cite{Bergshoeff:2014jla, Casalbuoni:2023bbh}.  We shall discuss the gauge invariance and the Carroll symmetry of the action in the next section.
 \par However, it is clear from the structure of the light-cone Minkowski metric that one cannot obtain an electric Carroll action from the Lorentzian action \eqref{PP_can LC}, since there is no $E^2$-term in the on-shell condition, which is a key feature of a electric Carroll particle.

 %%%%%%%%%%%%%%%%%%%%%%%%%%%%%%%%%%%%%%%%%%%%%%%%%%%%
\subsection{From deformed light-cone action}
In order to get the magnetic as well as electric Carroll theories, we need a more general action that is invariant under one of the Bargmann subgroups $\mathfrak{b}_{\pm}$, but not under the full Poincar\'e group $\mathfrak{p}$. We will therefore explicitly break Poincar\'e invariance down to one of its Bargmann subgroups, for which we need to choose a preferred null direction. Without loss of generality, we choose to work with the null vector $\mathfrak{n} = \del_+$, and hence the relevant subgroup for our purposes is $\mathfrak{b}_-$.
\par
To this end, we add to the action a term of the form $\mathfrak{n}^\mu \mathfrak{n}^\nu p_\mu \dot{x}_\nu$, which violates invariance under the Lorentz transformations $M_{i-}$ and $M_{+-}$ within $\mathfrak{p}$. We thus end up with a more general action invariant only under the subgroup $\mathfrak{b}_-$. In other words, we replace the light-cone Minkowski metric $\eta_{\mu \nu}$ with a general Bargmann metric of the form
\be
G_{\mu \nu} =\begin{pmatrix}
0&-1 & 0 \\
-1&-\alpha& 0 \\
0 & 0 & \delta_{ij}
\end{pmatrix} \,.
\ee
where $\alpha$ is a constant parameter. 
\par
As we will discuss below, the parameter $\alpha$ may be thought of as a relative coupling between the electric and magnetic parts~\cite{Ciambelli:2023xqk}, since the electric Carroll sector will arise from the new term $\alpha\, \mathfrak{n}^\mu \mathfrak{n}^\nu p_\mu \dot{x}_\nu$.  But it is more natural to view $\alpha$ as a deformation parameter that implements a change from the Lorentzian light-cone coordinates $(x^+,x^-, x^i)$ to the Bargmann light-cone coordinates $(x^+_\alpha, x^-_\alpha, x^i_\alpha)$
\bea \label{Lenz-coord}
x^+_{\alpha} = x^+ + \frac{\alpha}{2}\, x^- \,,  \quad x^-_{\alpha} = x^- \,,  \quad x^i_\alpha = x^i \,.
\eea
Such deformations have, in fact, been explored in the context of light-front quantization, mostly for theories in $1+1$ dimensions~\cite{Lenz:1991sa, Burkardt:1995ct}. We shall return to this point in the concluding section, but for the purposes of this paper, it is sufficient to treat the $\alpha$-term in the action as a deformation that we introduce by hand.
\par We, therefore, consider the Bargmann particle action
\be
\mathcal S^{Barg} [x^\mu, p_\mu, e] = \int d\tau [p_\mu \dot{x}^\mu - \frac{e}{2}  (G^{\mu \nu}p_\mu p_\nu +m^2)]\,,
\ee
which, in the deformed light-cone coordinates, explicitly reads
\be
\mathcal S^{Barg} [x^\mu, p_\mu, e] = \int d\tau [ p_+ \dot{x}^+ + p_- \dot{x}^- +p_i \dot{x}^i - \frac{e}{2} (-\alpha p^2_+ -2 p_- p_+ + p^ip_i + m^2)]\,.
\ee

We shall now perform the null reduction as a limiting procedure such that both the electric and magnetic Carroll particle actions follow from the Bargmann action above. In going from the Bargmann to the Carroll case, we restrict the worldline of the particle to $x^-(\tau) = x^-_0$. To this end, we impose $\dot{x}^-(\tau) = \varepsilon$, where $\varepsilon$ is a small parameter. We also relabel $p- = \chi / e$, as before.
\par
In the limit $\varepsilon \to 0$, the Bargmann particle action in $(d+1)$ dimensions reduces to a Carroll action in one lower dimension 
\be
\lim_{\varepsilon \to 0} {}^{(d+1)}\mathcal S^{Barg} [x^\mu, p_\mu, e]\quad  \longrightarrow \quad{}^{(d)}S^{Carr} [x^a, p_a, e]\,,
\ee
where the $x^a= \{x^+, x^i\}$ denotes the local coordinates in the Carrollian spacetime.
\par
We now wish to rescale the phase space variables with appropriate factors of $\varepsilon$ in a way that preserves the form of the kinetic term or the symplectic 2-form, \ie\,,
\be
{}^{(d+1)}\Omega_{Barg} = dp_\mu \wedge dx^\mu \quad \xrightarrow{\text{$\varepsilon \to 0$}}\quad  {}^{(d)}\Omega_{Carr} = dp_a \wedge dx^a\,,
\ee
and leads to the correct on-shell constraint for either the electric or magnetic Carroll case. It is clear from the expression above that there are two possible rescalings preserving the form of $\Omega$. In one case, we simply map the Bargmann momenta and coordinates $\{p_{+}, p_i, x^{+}, x^i\}$ to the Carrollian ones $\{p_a, x^a\}$ without any factor of $\varepsilon$. Another choice is to map the momenta and coordinates with inverse factors of $\varepsilon$. We shall explain below how these two choices lead to the magnetic and electric Carroll particle respectively.
\vskip 0.4cm
\ndt
\textit{Magnetic case:}
\vskip 0.2cm
\ndt
We first consider the following rescaling
\be
p_+ \to -E\,, \quad p_i \to p_i \,, \quad x^+ \to x^+\,, \quad x^i \to x^i\,, \quad \alpha \to \varepsilon \alpha\,,
\ee
along with $p_- \to \chi/e$. In the limit $\varepsilon \to 0$, the symplectic 2-form takes the form 
\be
{}^{(d)}\Omega_{Carr} = -dE \wedge dx^+ + dp_i \wedge dx^i\,.
\ee
The Bargmann action then reduces to a Carrollian action of the magnetic type
\be
{}^{(d)}\mathcal S_{mag} [x^+, x^i, E, p^i, \chi, e] = \int d\tau [-E \dot{x}^+ + p_i x^i - \frac{e}{2} (p^i p_i + m^2) + \chi E]\,.
\ee
The canonical generator of the Carroll transformations is given by
\be
G_C = (b_ix^i + a^+) E + (b_ix^i + a^+) p_i\,.
\ee
The magnetic phase space variables transform as follows
\bea
\delta_C x^+ = b_ix^i + a^+ \,, && \delta_C x^i = \omega^i_j x^j +a^i \,, \\
\delta_C E = 0\,, && \delta_C p_i = \omega^k_i P_k\,.
\eea
We see that the energy $E$ of the magnetic particle is invariant under Carroll transformations. 
\par In order to ensure the invariance of the action under Carroll boosts $b_i$, we must also transform the Lagrange multiplier $\chi$
\be
\delta_b \chi = e\, b^i p_i\,.
\ee
The invariance of the action under the rest of the Carroll transformations follows trivially. 
\par The time reparameterizations are generated by the first-class constraints in the theory
\be
G_{\lambda, \sigma} = \int d\tau \left[ \frac{1}{2} \lambda (\tau) (p^i p_i+ m^2) + \dot{\lambda}(\tau) \pi_e + \sigma (\tau) E\right]
\ee
Under these reparameterizations, the coordinates and Lagrange multipliers transform as
\be
\delta_{\lambda, \sigma} e = \dot{\lambda}\,, \quad \delta_{\lambda, \sigma} x^+ = \sigma\,, \quad \delta_{\lambda, \sigma} x^i = \lambda p^i \,, \quad  \delta_{\lambda, \sigma} \chi = \dot{\sigma}\,,
\ee
while the other variables are gauge-invariant. The transformation of $\chi$ is crucial to ensure the invariance of the magnetic Carroll action.

\newpage
\ndt
\textit{Electric case:}
\vskip 0.2cm
\ndt
We now consider the other case where the canonical variables are rescaled as
\be
p_+ \to -\varepsilon E\,, \quad p_i \to \varepsilon p_i \,, \quad x^+ \to \frac{x^+}{\varepsilon}\,, \quad x^i \to \frac{x^i}{\varepsilon}\,, \quad \alpha \to \frac{\alpha}{\varepsilon^2}\,,
\ee
In the limit $\varepsilon \to 0$, the symplectic 2-form again reduces to ${}^{(d)}\Omega_{Carr}$. The Bargmann action now yields the canonical action of the electric Carroll particle in $d$ dimensions
\be
{}^{(d)}\mathcal S_{elec} [x^+, x^i, E, p^i, e] = \int d\tau \left[-E \dot{x}^+ + p_i x^i - \frac{e}{2} (-\alpha^2 E^2 + m^2) \right] \,.
\ee
As in the magnetic case, the Carroll transformations are generated by 
\be
G_C = (b_ix^i + a^+) E + (b_ix^i + a^+) p_i \,,
\ee
with the electric phase space variables transforming as follows
\bea
\delta_C x^+ = b_ix^i + a^+ \,, && \delta_C x^i = \omega^i_j x^j +a^i \,, \\
\delta_C E = 0\,, && \delta_C p_i = \omega^k_i P_k\,.
\eea
The time reparameterization generator simply reads
\be
G_{\lambda} = \int d\tau \left[ \frac{1}{2} \lambda (\tau) (-\alpha^2 E^2+ m^2) + \dot{\lambda}(\tau) \pi_e\right]\,.
\ee
which generates non-trivial transformations only on the einbein $e$ and the $x^+$ coordinate
\be
\delta_{\lambda, \sigma} e = \dot{\lambda}\,, \quad \delta_{\lambda, \sigma} x^+ =\alpha^2 \lambda E\,.
\ee
It is straightforward to check that the electric Carroll action is invariant under these reparameterizations.
\par
By rescaling $\alpha \to \varepsilon \alpha$ in the magnetic case, one could argue that we are effectively removing the $\alpha$-deformation of the light-cone coordinates by hand. This interpretation is indeed correct, since one can already obtain the magnetic sector directly from the light-cone action. However, in the electric case, it is crucial that $\alpha \neq 0$ $-$ the electric Carroll sector is accessible via the null reduction procedure \textit{only} in the $\alpha$-deformed case.
\par
We shall see in the following sections that this is the key difference between the magnetic and electric sectors in this approach. Another way to put this is that for $\alpha = 0$ in the magnetic case, one obtains an additional constraint in the Hamiltonian theory, whereas the electric case, with $\alpha \neq 0$, is devoid of this extra constraint. This point was nicely demonstrated in~\cite{Bergshoeff:2022qkx}, where it was shown that the magnetic and electric sectors of Carroll theories differ precisely by the presence or absence of certain constraints in the theory.

\subsection{Carrollian field theories from null reduction}
We shall now shift our attention from point-particle actions to theories involving bosonic fields, with or without gauge symmetry. In this subsection, we describe a general procedure to restrict a Lorentzian field theory onto a null hyperplane such that the resulting action possesses Carrollian symmetry. The light front itself can be described as an embedding into light-cone Minkowski spacetime, as discussed in Appendix~\ref{Carroll-mani}. It is therefore possible to derive all relevant objects geometrically from their Lorentzian counterparts and construct the corresponding Carrollian theories using the intrinsic null geometry. However, here we employ a simpler trick that allows us to directly extract the Carrollian action from the higher-dimensional parent action~\cite{Chen:2023pqf}
\par
We consider a Hamiltonian action ${}^{(d+1)}\mathcal S$ in light-cone Minkowski spacetime for a theory involving a  collection of fields $\{\Phi\}$, their conjugate momenta $\{\Pi\}$ and possibly some Lagrange multipliers $\{\lambda \}$ to allow for general Hamiltonian dynamics~\footnote{We shall use the notation $\int d^\perp x$ and $\delta^\perp (x-y)$  as a shorthand for $\int dx^1 dx^2... dx^{(d-1)}$ and $\delta^{(d-1)}(x-y)$.}
\be \label{GenS}
{}^{(d+1)}\mathcal S[\Phi, \Pi, \lambda] = \int dx^+ dx^- d^\perp x\ {}^{(d+1)} \mathcal L_H [\Phi, \Pi, \lambda]\,.
\ee
The canonical Poisson brackets and the symplectic 2-form are given by
\be
\{ \Pi (x), \Phi (y) \} =\delta (x^- -y^-) \delta^\perp(x-y)\,, \quad  {}^{(d+1)} \Omega_{\varepsilon}    \int dx^- d^\perp x \, \delta_\varepsilon (x^- -x^-_0) \,d_V \Pi \wedge d_V \Phi\,.
\ee
\par
\begin{center}
\begin{figure}[h]
\centering
\includegraphics[width=2.5in]{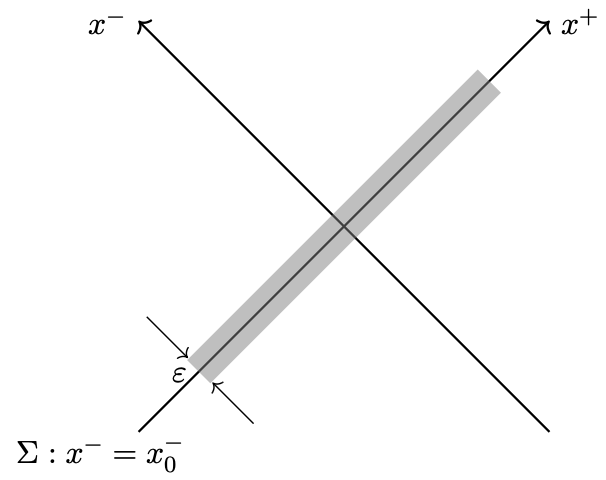}
%\begin{tikzpicture}
%  % Axes
%  \draw[->, line width=0.8pt] (-2.5,-2.5) -- (2.5,2.5) node[anchor=west] {\( x^+ \)};
%  \draw[->, line width=0.8pt] (2.5,-2.5) -- (-2.5,2.5) node[anchor=east] {\( x^- \)};
%  \draw[->, line width=0.5pt] (-2.4,-1.1) -- (-1.9,-1.6);
%  \draw[->, line width=0.5pt] (-1.1,-2.4)--(-1.6,-1.9) ;
%  \fill[gray,opacity=0.5,rotate around={135:(0,0)}] (-0.2,-2.5) rectangle (0.2,2.5);
%
%  % Event E on the strip
%  \node at (-2,-1.8) {$\varepsilon$ };
%
%  % Arrow for x^- = x^-_0
% % \draw[->,thick] (-1.5,-1.5) -- (-1.1,-1.1);
%  \node[anchor=north east] at (-1.8,-2.4) {\(\Sigma:  x^- = x_0^- \)};
%\end{tikzpicture}
  \caption{Null reduction to hypersurface $\Sigma: x^- = x^-_0$}
\label{fig:hypersurface}
\end{figure}
\end{center}

 
Now we wish to derive Carrollian actions by restricting the Lorentzian theory to the null hypersurface $\Sigma$. To evaluate the effective action restricted to any $x^- = x_0^- $ hypersurface, we introduce a Dirac delta distribution with a smearing function~\cite{Chen:2023pqf}
\bea
\delta_\varepsilon (x^- -x^-_0) &=& 
\begin{dcases}
      \frac{1}{\varepsilon} \,, &x^-_0 - \frac{\varepsilon}{2} < x^- < x^-_0 + \frac{\varepsilon}{2}  \\
0 \,,& \text{otherwise} 
    \end{dcases}
\eea
such that 
\be
\lim_{\varepsilon \to 0} \delta_\varepsilon (x^- -x^-_0) = \delta (x^- -x^-_0) \,. 
\ee
We, therefore, restrict the domain of $x^-$ to a finite interval of width $\varepsilon$ centered at $x^- = x^-_0$, thereby introducing a cutoff scale in the theory, as shown in Figure~\ref{fig:hypersurface}. 
\par Let us illustrate this using an arbitrary function $f$ of $x^-$. Inserting the delta function distribution into the integral given below
\be
\int_{-\infty}^{\infty} dx^- f(x^-)\quad \longrightarrow \quad \int_{-\infty}^{\infty} dx^-\delta_\varepsilon (x^- -x^-_0) f(x^-) \,,
\ee
we can isolate the value of  $f(x^-)$ in the neighborhood of $x^- = x^-_0$
\be
\quad \lim_{\varepsilon \to 0}\int_{-\infty}^{\infty} dx^-\delta_\varepsilon (x^- -x^-_0) f(x^-) ~=~ \lim_{\varepsilon \to 0} \frac{1}{\varepsilon} \int_{x^-_0 - \frac{\varepsilon}{2}}^{x^-_0 - \frac{\varepsilon}{2}}dx^- f(x^-) ~=~ f(x^-_0)\,.
\ee
\par
Going back to \eqref{GenS}, we get a family of actions $\mathcal S_\varepsilon$ by inserting the smearing function, such that in the limit $\varepsilon \to 0$, we obtain a Carrollian action
 \bea
{}^{(d)}\mathcal S_{Carr} ~\equiv ~\lim_{\varepsilon \to 0}  {}^{(d+1)}\mathcal S_{\varepsilon}   &=&\lim_{\varepsilon \to 0}   \int dx^+   \int d^\perp x \int dx^- \delta_\varepsilon (x^- -x^-_0)  {}^{(d+1)} \mathcal L_H \,.
 \eea
 
In order to pick out the finite, non-zero piece in the integral, we must specify the behavior of the ${(d+1)}$-dimensional fields around $x^- = x^-_0$.  This ensures that in the limit $ \varepsilon \to 0$, the Carrollian action so obtained is a functional of the Carrollian fields. We shall see that there are two allowed behaviors of the higher-dimensional fields in the vicinity of a constant $x^-$ plane, that preserve the canonical structure upon null reduction
\be
 {}^{(d)}\Omega_{Carr} ~\equiv ~ \lim_{\varepsilon \to 0} {}^{(d+1)} \Omega_{\varepsilon}\,.
\ee
These two possibilities lead to either the magnetic Carroll or electric Carroll theory in one lower dimension.
\par
In the case of gauge theories, we must also impose an additional condition on the gauge fields: the $d$-dimensional Carrollian fields ${\Phi_{\mu_1 \mu_2 \dots \mu_p}}$ must not have any components along $\mathfrak{m}^\mu$, which is orthogonal to the null plane $\Sigma$, \ie\,,
\be
\Phi \cdot \mathfrak m= {\Phi_{- \mu_2...\mu_p}} =0 \,.
\ee
This condition amounts to setting to zero the `minus' components $A_-$ of a vector field, or the field strength $F_{- \mu_{2} \dots \mu_{p}}$ for a $p$-form field, upon performing the null reduction.

\section{Scalars}
We begin with the case of scalar fields in Minkowski spacetime. After briefly describing the canonical analysis of the light-cone scalar field theory, we show how one can derive a magnetic Carrollian action directly from the Lorentzian action. However, as in the case of Carroll particles, to obtain both the electric and magnetic Carroll sectors, we must suitably deform the Lorentzian light-cone action, as discussed in the following subsection.
\subsection{From Lorentzian light-cone action }
Let us consider the Lorentzian action for massless scalar fields in light-cone coordinates
\be \label{LC-Scalar}
\mathcal S =  \int dx^+ dx^- d^\perp x\ \mathcal L^{Lor}  \,,
\ee
where the Lagrangian density reads
\be 
 \mathcal L^{Lor} = -\frac{1}{2} \eta^{\mu \nu}_{lc} \del_\mu \phi \del_\nu \phi ~=~ \del_- \phi \del_+ \phi - \frac{1}{2} \del_i \phi \del^i \phi\,.
\ee  This action is invariant under the light-cone Poincar\'e transformations listed in \eqref{Poincare}. We define the conjugate momenta with respect to the (Carrollian) time $x^+$
\be \label{Lor-pi}
\pi = \frac{\delta \mathcal L^{Lor}}{\delta(\del_+ \phi)} = \del_- \phi \,.
\ee
The canonical Poisson brackets are
\bea \label{PB} 
\{ \phi (x), \pi (y) \} \Big|_{x^+= y^+} &=& \delta(x^- - y^-) \delta^\perp (x-y) \,.
\eea
Since $\pi$ does not involve any time derivatives, we get a primary constraint
\be
\chi= \pi - \del_- \phi \,.
\ee
This primary constraint is second-class, which further allow us to go to a reduced phase space description~\cite{Steinhardt:1979it}. The detailed canonical analysis of light-cone scalar field theory is given in Appendix \ref{scalar-canonical}. Here, we only focus on the Lorentzian scalar field action in the Hamiltonian form.
\par
The canonical Hamiltonian density involves spatial gradients but does not contain any time derivatives or conjugate momenta
\be \label{LC-Ham}
\mathcal H_C = \pi \del_+ \phi - \mathcal L^{Lor} = \frac{1}{2} \del_i \phi \del^i \phi \,.
\ee
Thus, the Hamiltonian action takes the form
\be  \label{Sextended}
\mathcal S_{E} [\phi, \pi, \lambda ] = \int dx^- dx^+  d^\perp x  \Big\{ \pi \del_+ \phi - \frac{1}{2} \del_i \phi \del^i \phi - \lambda ( \pi -\del_- \phi ) \Big\} \,,
\ee
where the constraint is imposed through the Lagrange multiplier $\lambda$.
 The canonical structure is given by 
\be
{}^{(d+1)}\Omega = \int dx^- d^\perp x \ d_V \pi \wedge d_V \phi \,,
 \ee
where $d_V$ denotes exterior derivative in phase space. If we eliminate $\pi$ by strongly imposing the constraint $\chi$, we obtain a symplectic form in a reduced phase space involving only $\phi$,
\be
{}^{(d+1)}\Omega_{red} = \int dx^- d^\perp x \ d_V (\del_- \phi) \wedge d_V \phi \,.
 \ee
In this case, the canonical (Dirac) bracket is non-local in $x^-$
 \be
 \{ \phi(x), \phi(y)\} = \frac{1}{2} \theta (x^--y^-) \delta^\perp (x-y) \,,
 \ee 
 where $\theta (x^--y^-)$ denotes the Heaviside step function. However, in order to derive Carroll theories from light-cone Lorentzian actions, we must localize the $x^-$ integral around a constant value. Therefore, we shall continue to use the Poisson brackets \eqref{PB} and implement the constraint through a Lagrange multiplier.
\par
%\subsection{Null reduction to Carroll actions}
We now wish to restrict the theory to the hypersurface $\Sigma$ by considering certain $x^-$ behaviour of the fields. Let us assume that on $\Sigma$ we have
\bea \label{mag-behave1}
\phi(x^+,x^-, x^i) \big|_\Sigma &=&  \phi_m (x^+, x^i) \,, \\
\pi (x^+, x^-, x^i) \big|_\Sigma  &=&  \pi_m (x^+, x^i) \,, \\
\del_- \phi(x^+,x^-, x^i) \big|_\Sigma &=&p_m (x^+, x^i)\,.
\eea
We strongly impose the constraint $\chi$ by setting ${\pi}_m =  p_m$. In  other words, going from $(d+1)$ dimensions to the $d$-dimensional Carrollian spacetime, we relabel the Lorentzian fields as
\be \label{mag-rescale}
\phi \to \phi_m \,, \quad \pi = \del_- \phi \to p_m\,.
\ee
which leads to a well-defined canonical 2-form in the $d$-dimensional theory
\be
{}^{(d)}\Omega^{mag} \equiv\lim_{\varepsilon \to 0}{}^{(d+1)}\Omega^{Lor}_\varepsilon ~=~ \int d^\perp x \ d_V p_m \wedge d_V \phi_m \,,
\ee
where $\phi_m(x^+, x^i)$ and $p_m(x^+, x^i)$ are now the phase space variables of the Carroll Hamiltonian theory. 
\par
As $\varepsilon \rightarrow 0$, the Lorentzian action \eqref{S-HamScalar} reduces to the magnetic Carroll action
\be \label{S-Carr}
 {}^{(d)}\mathcal S^{mag} [\phi_m, p_m] =\int dx^+ d^{d-1} x \left( p_m \del_+ \phi_m -\mathcal H^{mag}\right)\,,
\ee
with the Carroll Hamiltonian and momentum densities
\be \label{mag-dens}
\mathcal H^{mag}= \frac{1}{2} \del_i \phi_m\del^i \phi_m \,, \quad \mathcal P_i^{mag} = p_m\del_i \phi_m\,.
\ee
The magnetic Carroll action so obtained is identical to the ones presented in~\cite{Henneaux:2021yzg, Chen:2023pqf}.

\par However, we note that it is not possible to derive the electric Carroll sector via null reduction from the Lorentzian light-cone action. This is because the canonical Hamiltonian \eqref{LC-Ham} is already of the magnetic form, \ie\,, it involves only spatial gradients even before we reduce the theory on one of the light fronts. In the next section, we discuss how to suitably \textit{deform} the Lorentzian light-cone action that will allow us to obtain both the electric and magnetic Carroll sectors in one lower dimension.

\subsection{From deformed light-cone action}
In order to obtain both the magnetic and electric Carroll theories, we must consider a more general action that is invariant under one of the Bargmann subgroups, but not under the full Poincar\'e group. Following the same procedure as in the previous section, we explicitly break Poincar\'e invariance by adding to the action~\eqref{LC-Scalar} a term of the form $\mathfrak{n}^\mu \mathfrak{n}^\nu \del\mu \phi \del_\nu \phi$. We thus arrive at a more general action, invariant only under the subgroup $\mathfrak{b}_-$.
\be \label{S-Barg-sc}
\mathcal{S}^{Barg}
= \int dx^+ dx^- d^\perp x \left( \frac{1}{2} \alpha \del_+ \phi \del_+ \phi + \del_- \phi \del_+ \phi - \frac{1}{2} \del_i \phi \del^i \phi \right),
\ee
where $\alpha$ is the deformation parameter as before and the factor of $1/2$ is included simply for convenience. 
\par
\par
Going to the Hamiltonian formulation , we find that the conjugate momenta 
\be
\pi  = \alpha \del_+ \phi + \del_- \phi \,,
\ee
are no longer constrained as in the Lorentzian case. The Hamiltonian action reads
\be \label{S-HamScalar}
\mathcal S^{Barg} ~=~   \int dx^+ dx^- d^\perp x \left( \pi \del_+ \phi -\mathcal H^{Barg}  \right)\,,
\ee
where the Hamiltonian density now involves both conjugate momenta and spatial gradients
\be
\mathcal H^{Barg}  = \frac{1}{2\alpha} (\pi - \del_- \phi)^2 + \frac{1}{2} \del_i \phi \del^i \phi\,.
\ee
The canonical brackets are given by 
\be
\{\pi (x), \phi (y) \} = \delta (x^- - y^-) \delta^\perp (x-y)\,,
\ee
while the symplectic form reads
\be
\quad \Omega^{Barg} = \int dx^- d^\perp x \ d_V \pi \wedge d_V \phi \,.
\ee
\par 
We now wish to restrict the theory to the hypersurface $\Sigma: x^- = x^-_0$ by inserting the smearing function $\delta_\varepsilon(x^- -x^-_0)$ into the Bargmann action. By assuming certain behaviour of the fields in the neighborhood of $x^- = x^-_0$, the Carroll action follows from
\be
{}^{(d)}\mathcal S^{Carr} \equiv \lim_{\varepsilon \to 0} {}^{(d+1)}\mathcal S^{Barg}_\varepsilon [\phi, \pi]\,.
\ee
Unlike in the Lorentzian case, there are now two possible choices for the $x^-$ behavior of the fields that preserve the structure of $\Omega^{Barg}$ as well as lead to a well-defined Carroll action. 
\vskip 0.3cm
\ndt
\textit{Magnetic case:}
\vskip 0.1cm
\ndt
To obtain the magnetic Carroll sector, we assume the same behaviour of the fields as in \eqref{mag-behave1}-\eqref{mag-rescale}. Therefore, we rescale the Bargmann fields and the parameter $\alpha$ as
\be \label{mag-rescale2}
\phi \to \phi_m \,, \quad \pi = \del_- \phi \to p_m\,, \quad \alpha \to \frac{\alpha}{\varepsilon^2} \,,
\ee
which leads to a well-defined canonical 2-form in the $d$-dimensional theory
\be
{}^{(d)}\Omega^{mag} \equiv\lim_{\varepsilon \to 0}{}^{(d+1)}\Omega^{Barg}_\varepsilon ~=~ \int d^{d-1}x \ d_V p_m \wedge d_V \phi_m \,.
\ee
As $\varepsilon \rightarrow 0$, the light-cone Bargmann action \eqref{S-HamScalar} reduces to the magnetic Carroll action
\be \label{S-Carr-mag}
 {}^{(d)}\mathcal S^{mag} [\phi_m, p_m] =\int dx^+ d^{d-1} x \left( p_m \del_+ \phi_m -\mathcal H^{mag}\right)\,,
\ee
with the Hamiltonian and momentum densities
\be \label{mag-dens}
\mathcal H^{mag}= \frac{1}{2} \del_i \phi_m\del^i \phi_m \,, \quad \mathcal P_i^{mag} = p_m\del_i \phi_m\,.
\ee
This action is the same as the one we obtained in the previous section from the null reduction of the Lorentzian light-cone action. 
\par The equations of motion derived from the magnetic Carroll action are
\bea
 \del_+ \phi_m&=&0 \,,\\
 -\del_+ p_m +\del_i \del^i  \phi_m&=&0 \,.
\eea
The first equation implies that a massless magnetic Carroll field cannot move, 
\be
\phi_m (x^+,x^i) = \phi_m^{(0)}(0, x^i)\,.
\ee
The second equation allows us to solve for $p^m$ as follows
\be
 p_m (x^+,x^i) =   x^+ \del_i \del^i\phi_m^{(0)} (0, x^i) +p_m^{(0)} (0,x^i)\,.
\ee
The initial data is therefore given by two arbitrary functions of the spatial coordinates, $\phi_m^{(0)} (0, x^i)$ and $p_m^{(0)} (0, x^i)$. We also note from the structure of the equations of motion that the momenta $p_m$ are not related to $\del_+ \phi_m$, just as in Lorentzian light-cone theories. However, unlike in the Lorentzian case, the momenta are not constrained through any primary constraints.
\vskip 0.3cm
\ndt
\textit{Electric case:}
\vskip 0.1cm
\ndt
When going from $(d+1)$ to $d$ dimensions, we can consider another rescaling of the fields
\be \label{elec-rescale}
\phi \to \varepsilon \phi_e, \quad \pi  \to \frac{p_e}{ \varepsilon}, \quad \alpha \to \frac{\alpha}{\varepsilon^2} \,,
\ee
which also leads to a well-defined symplectic 2-form
\be \Omega^{elec} \equiv\lim_{\varepsilon \to 0}{}^{(d+1)}\Omega^{Barg}_\varepsilon ~=~  \int d^{d-1}x\, d_V p_e \wedge d_V \phi_e  \,.
\ee
In the limit $\varepsilon \to 0$, we obtain the electric Carroll action from the Bargmann action \eqref{S-HamScalar}
\be \label{S-Carr-elec-sc}
 {}^{(d)}\mathcal S^{elec} [\phi_e, p_e]  = \int dx^+d^{d-1}x \left(p_e \del_+ \phi_e - \mathcal{H}^{elec}\right)  \,,
\ee
with the Hamiltonian and momentum densities
\be \label{elec-dens}
\mathcal{H}^{elec} =  \frac{1}{2\alpha}p^2_e\,, \quad \mathcal{P}_i^{elec} = p_e \del_i \phi_e\,.
\ee
Note that in the previous section, one could formally consider rescaling the fields of the Lorentzian theory in the same way as in \eqref{elec-rescale}. While this rescaling preserves the canonical structure, the Hamiltonian density $-$ lacking any terms depending on $\pi$ $-$ vanishes in the limit ${\varepsilon \to 0}$ . As a result, it is not possible to obtain a well-defined electric Carroll action directly from the Lorentzian theory.
\par
The equations of motion derived from the electric Carroll action are 
\bea \label{eom-elec}
\del_+p_e = 0 \,, \quad  \alpha \del_+ \phi_m =p_e   \,.
\eea
The first equation implies that $p_e$ is constant in time
\be
p_e (x^+, x^i) = p_e^{(0)}  (0, x^i)\,,
\ee
and the second equation yields
\be
\phi_e (x^+,x^i) = \frac{1}{\alpha} p_e^{(0)} (0, x^i) x^+ + \phi_e^{(0)}  (0, x^i)\,.
\ee
Therefore, the initial data is given by two arbitrary functions of the spatial coordinates, $\phi_e^{(0)} (0, x^i)$ and $p_e^{(0)}(0, x^i)$, as in the magnetic case.
\par 
From the second equation of motion in \eqref{eom-elec}, we observe that the electric Carroll momenta $p_m$ are directly related to the velocities $\partial_+ \phi_e$. This feature further emphasizes why the electric Carroll theory cannot be obtained directly from the Lorentzian action. In the Lorentzian light-cone phase space, the momenta are always constrained variables \eqref{Lor-pi} and are not related to the velocities. To obtain the electric sector, it is therefore essential to deform the theory to a Bargmann-invariant action. 
\par
\begin{figure}
    \centering
    \includegraphics[width=5in]{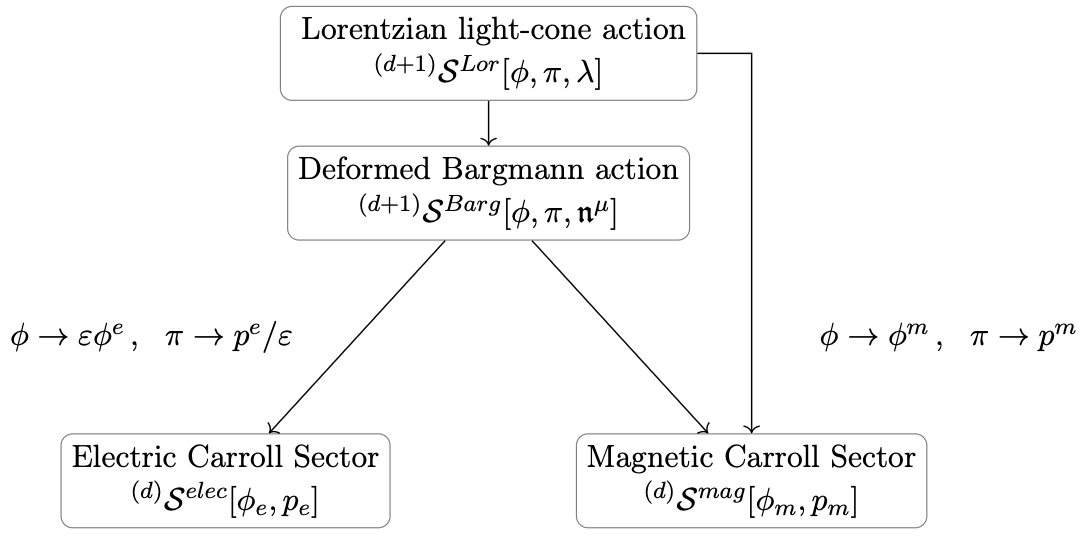}
%\begin{tikzpicture}
%\node (Rel LC2) [forces, yshift= -2.5cm] {\begin{varwidth}{20em} \begin{center}
%    \ Lorentzian light-cone action  \\${}^{(d+1)}\mathcal S^{Lor}[\phi,\pi, \lambda]$
%\end{center} \end{varwidth}};
%\node (Barg LC) [forces, below of = Rel LC2, yshift= -0.7cm] {\begin{varwidth}{20em} \begin{center}
%    Deformed Bargmann action \\ ${}^{(d+1)}\mathcal S^{Barg}[\phi,\pi, \mathfrak n^\mu]$
%\end{center}\end{varwidth}};
%\node (Electric2) [forces, below of =Barg LC , xshift= -3.2cm, yshift=-2.5cm] {\begin{varwidth}{20em} \begin{center} Electric Carroll Sector \\ $ {}^{(d)}\mathcal S^{elec} [\phi_e, p_e] $ \end{center}\end{varwidth}};
%\node (Magnetic2) [forces, below of = Barg LC, xshift= 3.2cm, yshift=-2.5cm] {\begin{varwidth}{20em}\begin{center}
%    Magnetic Carroll Sector \\  $ {}^{(d)}\mathcal S^{mag} [\phi_m, p_m] $
%\end{center}\end{varwidth}};
%\draw [->, line width=0.5pt, black] (Barg LC) --  (Electric2) node[midway, xshift= -2.5cm]{$ \phi \to \varepsilon \phi^e\,, \ \ \pi \to p^e/ \varepsilon $};
%\draw [->,line width=0.5pt, black] (Barg LC) --  (Magnetic2) node[midway, xshift = 4cm ]{$\phi \to \phi^m\,, \ \  \pi \to p^m$};
%\draw [->,line width=0.5pt, black] (Rel LC2) --  (Barg LC);
%\draw [->,line width=0.5pt, black] (Rel LC2) -| (Magnetic2);
%\end{tikzpicture}
    \caption{Two routes for deriving Carroll actions from Lorentzian light-cone actions}
    \label{fig:Carroll}
\end{figure}

Accordingly, we have presented two distinct approaches for deriving Carrollian field theories from a Lorentzian light-cone action: one that yields only the magnetic sector, and another one relying on a deformation of the Lorentzian action that leads to both electric and magnetic Carroll theories. This distinction is depicted schematically in Figure~\ref{fig:Carroll}.
\par
We can also include a mass term of the form $m^2 \phi^2$ in both the Lorentzian or the Bargmann-invariant action. In the magnetic sector, this term automatically survives the null reduction. However, in the electric sector, the mass parameter $m$ must be rescaled by a factor of $1/\varepsilon$ in order to retain the mass term in the $\varepsilon \to 0$ limit.
%\begin{center}
%\end{center}
\subsection{Carroll invariance and canonical generators}
In this section, we highlight certain properties that these Carroll theories share with the Carroll Hamiltonian theories obtained from the group contraction method (the $c \to 0$ limit of Poincar\'e theories) in~\cite{Henneaux:2021yzg}.
\par 
As described in Appendix \ref{Carroll-mani}, the Carrollian structure of the null  hypersurface $\Sigma$, given by the degenerate metric $h_{ab}$ and the null vector $\mathfrak n^a$, are preserved by the Carroll transformations
\be \label{Carroll}
\xi^+ = b_i x^i+ a^+\,, \quad \xi^i = \omega^i_j x^j + a^i\,.
\ee
where $b^i$ are the Carroll boosts. The canonical generator for these transformations reads 
\be \label{G-Carr}
G [{\xi^+, \xi^i}] = \int d^\perp x \ (\xi^+ \mathcal H^{C} + \xi^i \mathcal P^{C}_i)
\ee
where the Hamiltonian and momentum densities are either the electric or magnetic type as specified in \eqref{elec-dens} and \eqref{mag-dens}. Here, we use the notation $\phi^C, p^C$ to collectively denote the Carrollian fields which could be either electric, $\phi_e, p_e$ or magnetic $\phi_m, p_m$, unless specified otherwise. The commutation relations among the energy and momentum densities are 
\bea \label{H-P Comm}
\left[\mathcal{H}^{C}(x),\, \mathcal{H}^{C}(y) \right] &=& 0  \label{HH} \,, \\
\left[\mathcal{H}^{C}(x),\, \mathcal P^{C}_i (y)\right] &= &  \mathcal H^{C}(y) \,  \, \del_i \delta^\perp (x-y) \,, \\
\left[ \mathcal P^{C}_i (x),\, \mathcal P^{C}_j(y) \right] &=& \frac{1}{2}  \big[ \del_j \delta^\perp (x-y) \mathcal P^C_i (y) + \del_i \delta^\perp (x-y) \mathcal P^C_j (x) \big].
\eea
From~\eqref{HH}, we observe that the commutator of two Hamiltonian densities vanishes, which is a characteristic feature of Carrollian dynamics~\cite{Henneaux:1979vn}.
\bea \label{Carr-Gen}
&&E = \int d^\perp x \, \mathcal H^{C}(x) \,, \quad P_i = \int  d^{d-1}x\, \mathcal P^{C}_i (x) \,, \\
&& M_{ij} = \int d^\perp x\, (x_i \mathcal P^{C}_j - x_j \mathcal P^{C}_i) \,,\\
&& K_i = \int d^\perp x \,(x^+ \mathcal P^{C}_i - x_i \mathcal H^{C})\,.
\eea
The non-vanishing commutators of the Carroll algebra are
\bea \label{Carr-alg}
\left[ K_i ,\, P_j\right] &=& \delta_{ij} E \,, \\
\left[ M_{ij} ,\, P_l \right] &=& \delta_{lj} P_i - \delta_{il} P_j \,, \\
\left[ M_{ij} ,\, K_l \right] &=& \delta_{lj} K_i - \delta_{il} K_j \,, \\
\left[ M_{ij}, M_{lm} \right] &=& \delta_{im} M_{jl} - \delta_{il} M_{jm} - \delta_{jm} M_{il} + \delta_{jl}M_{im} \,.
\eea
One can easily verify that the corresponding field transformations generated by \eqref{G-Carr},
\be
\delta_{\xi^+, \xi^i} \phi^C =\{\phi^C, G[{\xi^+, \xi^i}]  \} \,, \quad \delta_{\xi^+, \xi^i} p^C   =\{p^C, G[{\xi^+, \xi^i}]  \}
\ee
render the Carrollian action \eqref{S-Carr} invariant. This fact, along with the algebra of constraints given by the set of equations \eqref{H-P Comm} constitute the necessary and sufficient condition for a Hamiltonian theory to be Carrollian~\cite{Henneaux:2021yzg}.
\par 
We now focus on the precise form of the transformation laws of the magnetic and electric Carroll fields under the isometries $(\xi^+, \xi^i)$. In the magnetic case, we find
\bea
\delta_{(\xi^+,\xi^i)} \phi_m &=& \xi^+ \del_+ \phi_m + \xi^i \del_i \phi_m ~=~ \xi^i \del_i \phi_m \,, \label{transfo1} \\
\delta_{(\xi^+,\xi^i)} p_m &=& \del^i (\xi^+ \del_i \phi_m) + \del_i (\xi^i p_m)\,,
\eea
where we have used the equation of motion for the last equality in \eqref{transfo1}. In particular, under Carroll boosts, $\xi = (b_i x^i, 0)$, the magnetic fields transform as
\be
\delta_b \phi_m = 0 \,, \quad \delta_b p_m = b_i \del^i \phi_m + b_j x^j \del_i \del^i \phi_m\,.
\ee
Similarly, in the electric case, the Carroll transformation laws are given by
\bea
\delta_{(\xi^+, \xi^i)} \phi_e &=& \xi^+ \del_+ \phi_e + \xi^i \del_i \phi_e ~=~ \xi^+  \frac{p_e}{\alpha} + \xi^i \del_i \phi_m \,,  \\
\delta_{(\xi^+, \xi^i)} p_e &=& \del^i (\xi^+ \del_i \phi_e) + \del_i (\xi^i p_e) = \del_i (\xi^i p_e) \,,
\eea
where we have used the equation of motion for the last equality in both the expressions. Under Carroll boosts, $\xi = (b_i x^i, 0)$, the transformation laws for the electric fields read
\be
\delta_b \phi_e = \frac{1}{\alpha}b_i x^i p_e  \,, \quad \delta_b p_e = 0\,.
\ee
These transformation laws are in complete agreement with the ones given in~\cite{Henneaux:2021yzg}. This further shows that although derived using a different technique, these Hamiltonian theories are the same as that obtained using the group contraction procedure.

\section{Gauge fields}
We now extend the procedure to various gauge theories. We begin with the example of electromagnetism, discussing the canonical analysis of light-cone gauge theories in both the Lorentzian and Bargmann cases. In particular, we highlight the role of gauge fixing in the null reduction method for obtaining Carrollian theories. We shall then generalize the procedure to non-Abelian gauge theories and $p$-form fields.
\subsection{Electromagnetism }
Let us consider the Maxwell action
\be
\mathcal{S}^{Lor} =  \int dx^+ dx^- d^\perp x\ \, \mathcal L_{EM}\,,
\ee
with Lorentzian light-cone Lagrangian
\be \label{LorL-EM}
\mathcal L_{EM} = -\frac{1}{4} F_{\mu \nu }F^{\mu \nu }  =\frac{1}{2} (F_{+-})^2 + F_{+i} F_{-i} - \frac{1}{4} F_{ij}F^{ij} \,,
\ee
where the Abelian field strength reads $F_{\mu \nu } = \del_\mu A_\nu -\del_\nu A_\mu$.
\par
In the Hamiltonian formulation, we find that some of the conjugate momenta are constrained, as in the case of scalars. The Hamiltonian analysis of light-cone electromagnetism is detailed in Appendix \ref{Maxwell-canonical}.
\par 
To obtain both magnetic and electric Carroll theories, we need to deform this action to a more general Bargmann action by adding a term of the form $\mathfrak n^\mu \mathfrak n^\nu\eta^{\rho \sigma}_{lc}  F_{\mu \rho }F_{\nu \sigma}$ to the Lorentzian Lagrangian \eqref{LorL-EM}
\be \label{S-Barg-EM}
\mathcal{S}^{Barg} =\int dx^+ dx^- d^\perp x\ \left\{\frac{1}{2} \alpha (F_{+i})^2 +\frac{1}{2} (F_{+-})^2 + F_{+i} F_{-i} - \frac{1}{4} F_{ij}F^{ij} \right\} \,.
\ee
The conjugate momenta are given by
\bea
\pi^+ = 0\,, \quad \pi^- = F_{+-} \,, \quad \pi^i = F_{-i} + \alpha F_{+i}\,.
\eea
The momenta $\pi^i$ are no longer constrained but explicitly depend on the velocities, as opposed to the Lorentzian case. The Hamiltonian density reads
\be
\mathcal{H}^{Barg} ~=~ \frac{1}{2} (\pi^-)^2 + \frac{1}{2\alpha} (\pi_i - F_{-i})^2 + \frac{1}{4} F_{ij}F^{ij} - A_{+}\mathcal G \,.
\ee
The first class constraint $\mathcal G= \del_- \pi^- + \del_i \pi^i$ defines the canonical generator
\be
{}^{(d+1)}G[\epsilon] = \int dx^- d^\perp x\ \epsilon(x) \mathcal G (x) \,,
\ee
that generates the gauge transformations on the $(d+1)$-dimensional fields
\be
\delta_\epsilon A_\mu = \del_\mu \epsilon\,, \quad \delta_\epsilon \pi^\mu =0\,. 
\ee
The Hamiltonian action for the Bargmann-invariant theory reads
\be \label{S-HamEM}
\mathcal{S}^{Barg}_H ~=~ \int dx^+ dx^- d^\perp x \left\{ \pi^i \del_+ A_i + \pi^- \del_+ A_- - \mathcal{H}^{Barg} \right\} \,.
\ee
We now restrict the theory to the light front $\Sigma$ using the null reduction method. As discussed earlier, we also need to ensure that the vector fields have no components off the hypersurface $\Sigma$. To achieve this, we set $A_-$ and its conjugate momentum $\pi^-$ to zero. This is, in fact, the light-cone gauge condition extensively used in quantum field theories in the light-cone formalism~\cite{Kogut:1969xa,Brodsky:1997de}.
\vskip 0.3cm
\ndt
\textit{Magnetic case:}
\vskip 0.1cm
\ndt 
To obtain the magnetic Carroll sector, we rescale the remaining components of the Maxwell fields and the momenta as follows
\be  \label{mag-spin1}
A_i \to A_i^m \,, \quad A_+ \to A_+^m\,, \quad \pi^i \to p^i_m\,, \quad \alpha \to \frac{\alpha}{\varepsilon^2} \,.
\ee
To simplify the notation, we drop the subscript $m$ on the magnetic Carroll fields. As $\varepsilon \rightarrow 0$, the light-cone Bargmann action \eqref{S-HamEM} reduces to the Carroll action
\be \label{S-Carr-mag-EM}
 {}^{(d)}\mathcal S^{mag} [A_i, p^i] =\int dx^+ d^\perp x \left( p^i \del_+ A_i -\mathcal H^{mag} - A_{+} \mathcal G_{Carr}\right)\,,
\ee
with the magnetic Carroll Hamiltonian and momentum densities
\be \label{mag-densEM}
\mathcal H^{mag}= \frac{1}{4} F_{ij}F^{ij}\,, \quad \mathcal P_i^{mag} = p^j\del_i A_j\,.
\ee
In Appendix \ref{Maxwell-canonical}, we show that the same action can be obtained directly from the Lorentzian action without any deformation.
The Carroll Gauss constraint reads $G_{Carr} = \del_k p^k $. The equations of motion obtained from the Hamiltonian action are 
\be
 \del_+ A_i - \del_k A_+=0 \,, \quad -\del_+ p^i +\del_j F^{ij}= 0 \,, \quad \del_k p^k = 0\,.
\ee
Using some simple redefinitions
\be
p^i \equiv E^i\,, \quad F^{ij} \equiv \frac{1}{2} \epsilon^{ijk}B_k\,,
\ee
we can recast these field equations into the familiar form of Maxwell equations for the magnetic case 
\be
\nabla \cdot\vec{E} = 0 \,,\quad  -\frac{\del \vec{E}}{\del x^+} + \nabla \times \vec{B} = 0\,, \quad \nabla \cdot\vec{B} = 0 \,, \quad \frac{\del \vec{B}}{\del x^+} =0\,,
\ee
where we have used the Bianchi identities for the last two equations.
\vskip 0.3cm
\ndt
\textit{Electric case:}
\vskip 0.1cm
\ndt 
To obtain the electric Carroll sector, we rescale the vector fields and the momenta as follows
\be \label{elec-spin1}
A_i \to \varepsilon A_i^e \,, \quad A_+ \to \varepsilon A_+^e\,, \quad \pi^i \to \frac{p^i_e}{\varepsilon}\,, \quad \alpha \to \frac{\alpha}{\varepsilon^2}\,,
\ee
For simplicity, we again drop the subscript $e$ on the electric Carroll fields. As $\varepsilon \rightarrow 0$, the light-cone Bargmann action \eqref{S-HamEM} reduces to the electric Carroll action
\be \label{S-Carr-elec-EM}
 {}^{(d)}\mathcal S^{elec} [A_i, p^i] =\int dx^+ d^\perp x \left( p^i \del_+ A_i -\mathcal H^{elec} - A_{+} \mathcal G_{Carr}\right)\,,
\ee
with the Carroll Hamiltonian and momentum densities
\be \label{elec-densEM}
\mathcal H^{elec}= \frac{1}{2\alpha} p^i p_i\,, \quad \mathcal P_i^{elec} = p^j\del_i A_j\,.
\ee
The equations of motion obtained from the Hamiltonian action are 
\be
 \del_+ A_i - \del_i A_+ - \frac{p_i}{\alpha} = 0 \,, \quad \del_+ p^i= 0 \,, \quad \del_k p^k = 0\,.
\ee
Again, using some simple redefinitions
\be
\frac{p^i}{\alpha} \equiv E^i\,, \quad F^{ij} \equiv \frac{1}{2} \epsilon^{ijk}B_k\,,
\ee
we can rewrite the equations of motion as the Maxwell equations for the electric case 
\be
\nabla \cdot\vec{E} = 0 \,, \quad \frac{\del \vec{E}}{\del x^+} =0\,,\quad  \frac{\del \vec{B}}{\del x^+} + \nabla \times \vec{E} = 0\,, \quad \nabla \cdot\vec{B} = 0 \,,
\ee
where we have used the Bianchi identities for the last two equations. This shows that the electric and magnetic Carroll theories so obtained agree with the ones presented in~\cite{Duval:2014uoa, Henneaux:2021yzg, Basu:2018dub}.
\par 
In both the electric and magnetic cases, the gauge transformations on the $d$-dimensional Carroll fields are now generated by 
\be
{}^{(d)}G_{Carr}[\epsilon ]= \int  d^\perp x\ \epsilon(x) \mathcal G_{Carr} (x)  =\int  d^\perp x\ \epsilon\, \del_k p^k_C \,,
\ee
where $p^k_C$ is either the electric or magnetic Carrollian momenta. The Carroll transformations are generated by $G[\xi^+, \xi^i]$, as mentioned in the previous section, with the Hamiltonian and momentum densities replaced by their Maxwell counterparts. One can check the Maxwell Hamiltonian and momentum densities in both electric and magnetic cases also satisfy the Carrollian hypersurface deformation algebra \eqref{H-P Comm}.

\subsection{Light-cone gauge}
One important reason for the effectiveness of the light-cone formalism is the convenient choice of gauge, which allows one to eliminate the redundant fields and express the theory entirely in terms of its physical degrees of freedom. Accordingly, we shall now consider the light-cone gauge-fixed formulation of electromagnetism~\cite{Kogut:1969xa} as our starting point for the null reduction procedure and show that the resulting Carroll actions are identical to those discussed in the previous section.
\par Starting from the usual Maxwell action, 
\be
\mathcal{S}^{Lor} =  \int dx^+ dx^- d^\perp x\ \, \mathcal L_{EM}\,,
\ee
we impose the light-cone gauge condition $A_- =0$, which modifies the form of the Largangian
\be
\mathcal L_{EM} = -\frac{1}{4} F_{\mu \nu }F^{\mu \nu }  =\frac{1}{2} (\del_-A_+)^2 + F_{+i} (\del_- A_i) - \frac{1}{4} F_{ij}F^{ij}\,. 
\ee
This action can be further simplified by solving one of the Maxwell equations that leads to a constraint
\be
\del_\mu F^{\mu +} = \del_-^2 A_+ -\del_- \del_k A_k = 0 \,.
\ee
Assuming that the fields fall-off at most as $\mathcal O(1)$ at large $x^-$, we can solve for $A_+$ in the above equation as follows
\be
A_+ (x^+, x^-, x^i) = \frac{\del_k}{\del_-}  A^k+ A_{+}^{(0)} (x^+, x^i) \,,
\ee
where $A_{+}^{(0)} (x^+, x^i)$ represents the boundary degree of freedom and $(\del_-)^{-1}$ is defined as in~\cite{Kogut:1969xa,Majumdar:2022fut} . Substituting this back to the Maxwell action yields the gauge-fixed on-shell action
\be
\mathcal{S}^{Lor} =  \int dx^+ dx^- d^\perp x\ \, \mathcal L^{Lor}_{on-shell}\,,
\ee
with the Lagrangian
\be \label{Lag-LCG}
 \mathcal L^{Lor}_{on-shell} =  \del_+ A^i \del_- A_i - \frac{1}{2} \del_j A^i \del^j A_i - \del_k A_{+}^{(0)} \del_- A_k\,.
\ee
We note that the last term involving the $A_{+}^{(0)}$ field is indeed a boundary term $\int dx^+ d^\perp x\, \del_k A_{+}^{(0)} A_k$, which accounts for the residual or large gauge transformations in the theory preserving the gauge condition $A_- =0$~\cite{Majumdar:2022fut}. But we keep it as a bulk term because here we are more interested in its role in the canonical analysis. 
\par
The conjugate momenta again lead to a first-class constraint $C_0 = \pi^+_{(0)}$ and a second-class constraint $C_i = \pi_i - \partial_- A_i$, whose form differs slightly from the ungauge-fixed case. We then perform the Dirac-Bergmann analysis of constraints and find that $C_0$ leads to an additional first-class Gauss-like constraint $\mathcal{G}_0 = \partial_k \pi^k$, which generates large gauge transformations. The Hamiltonian density takes the form
\be
\mathcal{H}_{on-shell} = \frac{1}{2} \del_j A^i \del^j A_i -  A_+^{(0)} \del_K \pi^k- \lambda^i (\pi_i -\del_- A_i)\,.
\ee
The gauge-fixed light-cone phase space consists of the fields $A_i$ and their momenta $\pi^i$, along with the boundary field $A_+^{(0)}$ and $\lambda^i$, appearing as Lagrange multipliers. The Hamiltonian action is given by
\be
\mathcal{S}_{on-shell} = \int dx^+ dx^- d^\perp x \left\{ \pi^i \del_+ A_i + \frac{1}{2}  \del_j A^i \del^j A_i -  A_+^{(0)}\mathcal G_0 - \lambda^i C_i\right\} \,.
\ee
Now we turn to the null reduction procedure, taking the on-shell gauge-fixed Lagrangian action as our starting point. To deform this action and render it Bargmann-invariant, we add the term $\mathfrak n^\mu \mathfrak n^\nu \del_\mu A_i \del_\nu A^i$ to the Lagrangian \eqref{Lag-LCG}.
\be
 \mathcal L^{Barg}_{on-shell} = \frac{1}{2} \alpha (\del_+ A_i)^2 +  \del_+ A^i \del_- A_i - \frac{1}{2} \del_j A^i \del^j A_i - \del_k A_{+}^{(0)} \del_- A_k \,.
\ee
The corresponding Hamiltonian density becomes
\be
\mathcal{H}^{Barg}_{on-shell} ~=~ \frac{1}{2} (\pi^-)^2 + \frac{1}{2\alpha} (\pi_i - \del_-A_i)^2 + \frac{1}{2} \del_j A^i \del^j A_i \,,
\ee
which no longer involves any second class constraints for the $\pi^i$. From the Bargmann-invariant Hamiltonian action
\be
\mathcal{S}^{Barg}_{on-shell} = \int dx^+ dx^- d^\perp x \left( \pi^i \del_+ A_i - \mathcal{H}^{Barg}_{on-shell} - A_{+}^{(0)}\mathcal G_0  \right)
\ee
we can derive Carrollian electric and magnetic theories following the same procedure as before. Mapping the $(d+1)$ dimensional fields to the $d$ dimensional Carroll fields as follows
\be  \label{mag-spin1}
A_i \to A_i^m \,, \quad A^{(0)}_+ \to A_+^m\,, \quad \pi^i \to p^i_m\,, \quad \alpha \to \frac{\alpha}{\varepsilon^2} \,,
\ee
gives the same magnetic Carroll action \eqref{S-Carr-mag-EM}, while the other map
\be \label{elec-spin1}
A_i \to \varepsilon A_i^e \,, \quad A^{(0)}_+ \to \varepsilon A_+^e\,, \quad \pi^i \to \frac{p^i_e}{\varepsilon}\,, \quad \alpha \to \frac{\alpha}{\varepsilon^2} \,,
\ee 
yields the electric Carroll action \eqref{S-Carr-elec-EM}. Note that $\mathcal{H}^{mag}$ may have a slightly different form, but it still contains only the magnetic part and involves only spatial gradients of $A_i$. One can readily check that first-class gauge constraint $\mathcal G_0$ is already of the form of the Carrollian gauge constraint, $\mathcal G_{Carr}$.
\par
The key point we wish to emphasize here is that the light-cone gauge condition $A_- = 0$ must be implemented at some point in the procedure in order to obtain the Carroll theories; however, the order in which it is applied does not matter. One could either start with the covariant Lorentzian action, deform it to the Bargmann action, and then impose the light-cone gauge condition \textit{while performing the null reduction}, as shown in the previous section . Alternatively, one could \textit{gauge-fix the Lorentzian action right away}, then deform it to a Bargmann-invariant action, and finally perform the null reduction, as described in this section. In both cases, one obtains the same Carroll theory in one lower dimension. Although the two procedures yield different phase spaces for the intermediate Bargmann theory, they reduce to the same Carrollian phase space after null reduction. We summarize this point with the help of a schematic diagram in Figure \ref{fig:LCgauge}.
\begin{center}
\begin{figure}
    \centering
  \includegraphics[width=5.3in]{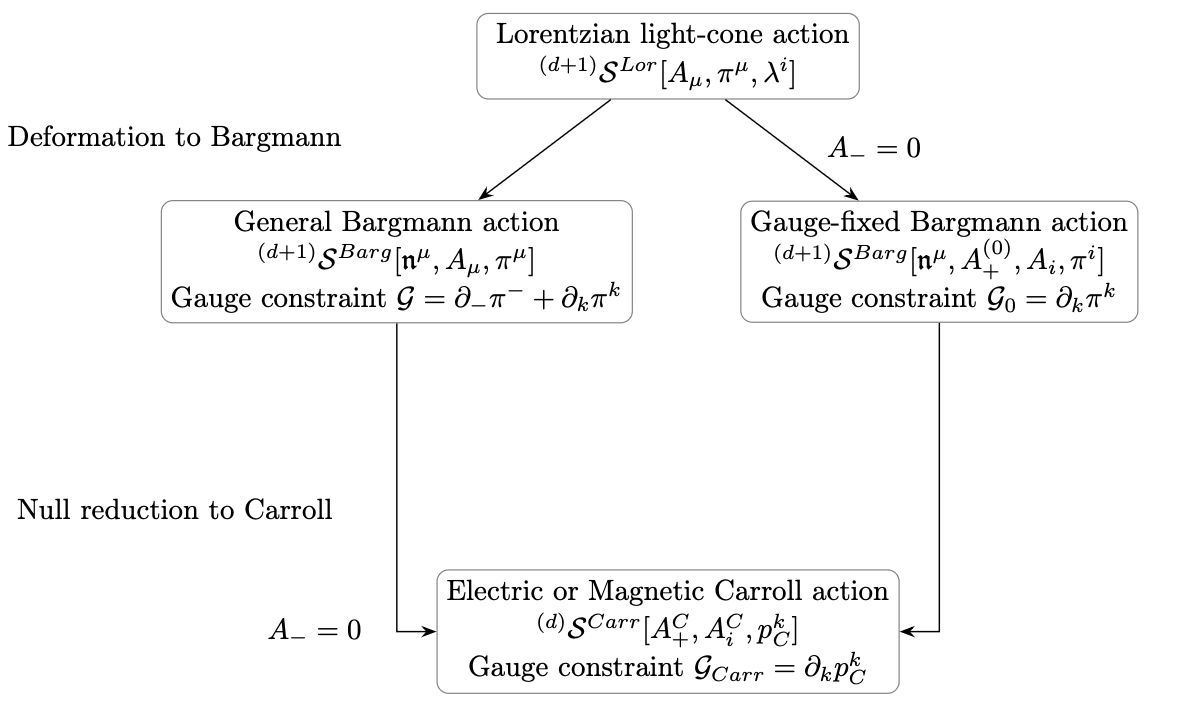}  
   \caption{Light-cone gauge condition for deriving Carrollian actions}
    \label{fig:LCgauge}
\end{figure}
\normalsize
\end{center}

\subsection{Yang-Mills theory}
We now turn to the case of Yang-Mills theory with the gauge group $SU(N)$. The field strength for the non-Abelian fields is given by
\be
F^A_{\mu \nu }= \del_\mu A^A_\nu -\del_\nu A^A_\mu -g f^{ABC} A^B_\mu A^C_\nu\,,
\ee
where the gauge indices $A, B, C$ run from $1$ to $N$. The generators of the $SU(N)$ Lie algebra satisfy the relations
\be
[T^A, T^B] = f^{ABC} T^C\,, \quad Tr(T^A T^B) = \frac{1}{2} \delta^{AB} \,.
\ee
The Lorentzian Yang-Mills Lagrangian takes the form
\be
\mathcal L^{Lor}_{YM} = - \frac{1}{4} Tr(F^{\mu \nu } F_{\mu \nu}) = \frac{1}{2} F_{+-}^A F^A_{+-} + F^A_{+i}F^A_{-i} -\frac{1}{4} F^A_{ij} F^{ijA} \,.
\ee
Going to the Hamiltonian formulation, we find that the conjugate momenta are
\bea
\pi^{+A} ~=~ 0\,, \quad 
\pi^{-A} ~=~ F^A_{-+}\,, \quad
\pi^{iA} ~=~ F^A_{-i} \,.
\eea
The theory has second-class primary constraints, as in the case of electromagnetism,
\be
C^{iA} = \pi^{iA} - F^A_{-i}\,,
\ee
as well as a first-class primary constraint 
\be
C^A = \pi^{+A}\,,
\ee
that further leads to the Gauss constraint in the Yang-Mills theory
\be
\mathcal G^A   = \del_- \pi^{-A} + g f^{ABC}  A_-^B\pi^{-C} +\del_k \pi^{kA} + g f^{ABC}  A_k^B\pi^{kC}= \mathcal D_- \pi^{-A} + \mathcal D_k \pi^{kA} \,,
\ee
where $\mathcal D_\mu$ denote the  covariant derivatives. The corresponding canonical generator
\be
G [\epsilon^A] = \int dx^- d^\perp x\ \epsilon^A (x) \mathcal G^A (x) \,,
\ee
 generates the gauge transformations on the $(d+1)$ dimensional non-Abelian vector fields
\be
\delta_\epsilon A^A_\mu = \mathcal D_\mu \epsilon^A\,, \quad \delta_\epsilon \pi^{\mu A} =0\,. 
\ee
The canonical Hamiltonian density has the same form as in electromagnetism with the spacetime derivatives $\del_\mu$ replaced by covariant derivatives $\mathcal D_\mu$. After going through the Dirac-Bergmann constraint analysis following Appendix \ref{Maxwell-canonical}, we finally arrive at the Hamiltonian density for Yang-Mills theory
\be
\mathcal H^{Lor}_{YM} = \frac{1}{2} (\pi^{-A})^2 + \frac{1}{4} F_{ij}^A F^{ijA} + \lambda^{iA} (\pi_k^A - F_{-k}^A) + A_{+}^A \mathcal G^A \,,
\ee
and the light-cone Yang-Mills action in the Hamiltonian form
\bea
\mathcal S_{YM}^{Lor} &=& \int dx^+ dx^- d^\perp x \left\{ \pi^{-A} \del_+ A_-^A + \pi^{iA} \del_+ A^A_i -\mathcal H^{Lor}_{YM} \right\} \,.
\eea
On restricting this action to a constant null plane, we can obtain a magnetic Carroll action in $d$ dimensions. But since we are interested in both the electric and magnetic sectors, we shall directly go to the Bargmann case.
\par 
We deform the Lagrangian by adding a term of the form $\mathfrak n^\mu \mathfrak n^\nu \eta^{\rho \sigma} F^A_{\mu \rho} F^A_{\nu \sigma}$. Therefore, the resulting Bargmann invariant action reads
\be
\mathcal S^{Barg}_{YM} = \int dx^+ dx^- d^\perp x  \mathcal L^{Barg}_{YM} \,,
\ee
with the Lagrangian density
\bea
\mathcal L^{Barg}_{YM} &=& \frac{1}{2}\alpha\, \mathfrak n^\mu \mathfrak n^\nu \eta^{\rho \sigma} F^A_{\mu \rho} F^A_{\nu \sigma} + \mathcal  L^{Lor}_{YM} \nn \\
&=&\frac{1}{2} \alpha\,  F_{+i}^A F^A_{+i} + \frac{1}{2} F_{+-}^A F^A_{+-} + F^A_{+i}F^A_{-i} -\frac{1}{4} F^A_{ij} F^{ijA} \,.
\eea
So far, we have focused on deriving Carrollian actions using the Hamiltonian formulation. This approach was primarily intended to highlight the nature of constraints in light-cone theories and to discuss how they differ in the Lorentzian and Bargmann-invariant cases. It also allowed us to verify that the resulting Hamiltonian actions satisfy the necessary and sufficient conditions for Carrollian field theories.
\par
We now shift focus to the Lagrangian formulation and derive Carrollian theories directly within this framework. Using Yang-Mills theory as an example, we illustrate how to obtain both electric and magnetic Carroll sectors from the higher-dimensional Lagrangian action $\mathcal S^{Barg}_{YM}$ by putting appropriate conditions on the components of the field strength. We shall then extend this Lagrangian approach to general $p$-form gauge fields in the following section.
\par
To obtain the Carroll actions, we insert the smearing function into the Yang-Mills Bargmann action and suitably rescale the fields in the limit $\varepsilon \to 0$ 
\be
\mathcal S^{Carr}_{YM} ~=~ \lim_{\varepsilon\to 0}   \int dx^+ dx^- d^\perp x  \ \delta_\varepsilon (x^- - x^-_0)\mathcal L^{Barg}_{YM} \,.
\ee
To project the vector fields onto the null hypersurface, we also set to zero all the minus components of the field strength, \ie\,, $F^A_{-\mu} = 0$. We now discuss two distinct rescalings that lead to the electric and magnetic Carroll sectors.
\vskip 0.3cm
\ndt
\textit{Magnetic case:}
\vskip 0.1cm
\ndt 
We assume the behaviour of the rest of the components of $F^A_{\mu \nu}$ around $x^- = x^-_0$ to be
\bea
F^A_{+i} (x^+,x^-,x^i)\Big|_\Sigma & =&0 \,, \\
F^A_{ij} (x^+,x^-,x^i)\Big|_\Sigma & =& \Tilde{F}^A_{ij}(x^+,x^i) \,.
\eea
In other words, we choose the following map between the $(d+1)$-dimensional fields and the $d$ dimensional Carrollian fields
\be \label{mag-rescaleYM}
F^A_{+i} \longrightarrow 0\,, \quad F^A_{ij} \longrightarrow \tilde{F}^A_{ij} \,.
\ee
The Bargmann action reduces to the magnetic Carroll Yang-Mills action~\cite{Duval:2014uoa}
\be \label{S-Carr-magYM}
 {}^{(d)}\mathcal S^{mag}  = -\frac{1}{4} \int dx^+ d^\perp  x \ \tilde{F}^A_{ij}\tilde{F}^{ijA} = -\frac{1}{2}\int dx^+ d^\perp  x\ B^{A}_k B^{kA} \,,
\ee
where $B^{A}_i = \epsilon_{ijk}\tilde{F}^{jkA}/2$ is the Carrollian magnetic vector in $d$ dimensions.
\vskip 0.3cm
\ndt
\textit{Electric case:}
\vskip 0.1cm
\ndt 
Alternatively, we can choose the $x^-$ behaviour to be
\bea
F^A_{+i} (x^+,x^-,x^i)\Big|_\Sigma & =&\tilde{F}^A_{+i} (x^+,x^i) \,, \\
F^A_{ij} (x^+,x^-,x^i)\Big|_\Sigma & =& 0\,.
\eea
Thus, we choose the following map for the rest of the components of $F^A_{\mu \nu}$
\be \label{elec-rescaleYM}
F^A_{+i} \longrightarrow \tilde{F}^A_{+i}\,, \quad F^A_{ij} \longrightarrow 0 \,.
\ee
This map reduces the Bargmann action to the electric Carroll action~\cite{Duval:2014uoa}
\be \label{S-Carr-elecYM}
 {}^{(d)}\mathcal S^{elec}  = \frac{1}{2} \alpha \int dx^+ d^\perp  x\ \tilde{F}^A_{+i}\tilde{F}^{A}_{+i} = \frac{1}{2}\alpha \int dx^+ d^\perp  x\ E_k^A E^{kA} \,,
\ee
where $E^A_i = \tilde{F}_{+i}$ is the Carrollian electric vector in $d$ dimensions.

\subsection{$p$-form fields}
Now, we can generalize this procedure to any $p$-form fields. We consider a $p$-form field in the $(d+1)$-dimensional Minkowski spacetime
\be A =\frac{1}{p!} A_{\mu_1...\mu_p} dx^{\mu_1} \wedge \cdots \wedge dx^{\mu_p}\,,
\ee
with the $(p+1)$-form field strength $F = dA$. The Lorentzian action for free $p$-form fields is given by
\be
\mathcal S^{Lor}_{p\text{-form}} = \int dx^+ dx^- d^\perp x \ \mathcal L^{Lor}_{p\text{-form}} \,, 
\ee
with the Lagrangian density
\be 
\mathcal L^{Lor}_{p\text{-form}}= - \frac{1}{2(p+1)!} \eta^{\mu_{1} \nu_{1}} \cdots \eta^{\mu_{p+1} \nu_{p+1}} F_{\mu_{1} \mu_{2} ... \mu_{p+1}}F_{\nu_{1} \nu_{2} ... \nu_{p+1}}\,.
\ee
To this Poincar\'e-invariant action, we add a term involving the null vector $\mathfrak n^\mu$ to get the Bargmann action
\be
\mathcal S^{Barg} = \int dx^+ dx^- d^\perp x\ \mathcal L^{Barg}_{p\text{-form}}\,,
\ee
with the Lagrangian density
\bea 
\mathcal L^{Barg}_{p\text{-form}} &=& - \frac{\alpha}{2(p+1)!} \, \mathfrak n^{\mu_1} \mathfrak n^{\nu_1} \eta^{\mu_2\nu_2} \ldots \eta^{\mu_p\nu_p} F_{\mu_1...\mu_{p+1}} F_{\nu_1...\nu_{p+1}} + \mathcal L^{Lor}_{p\text{-form}}\,.
\eea
We can now restrict the theory to a constant null hypersurface by suitably rescaling the fields. First, we set the `minus' components to zero, \ie\,, $ F_{-\mu_1...\mu_{p}} =0$.
\vskip 0.3cm
\ndt
\textit{Magnetic case:}
\vskip 0.1cm
\ndt 
For the magnetic sector, we assume that the $(d+1)$-dimensional fields are mapped to the Carrollian fields as follows
\be \label{mag-rescale-p}
F_{+i_1...i_{p}} [A] \longrightarrow  0\,,  \quad F_{i_1...i_{p+1}}  [A]\longrightarrow  \tilde{F}_{i_1...i_{p+1}} [A_m]\,.
\ee
The Bargmann action then reduces to magnetic Carroll action for free $p$-form fields in the Lagrangian form
\be \label{S-Carr-mag-p}
 {}^{(d)}\mathcal S^{mag} = \lim_{\varepsilon \to 0} \mathcal S^{Barg}_{p\text{-form}} =- \frac{1}{2(p+1)!} \int dx^+ d^\perp  x\,( \tilde{F}_{i_1...i_{p+1}})^2  \,.
\ee

\vskip 0.3cm
\ndt
\textit{Electric case:}
\vskip 0.1cm
\ndt 
If we consider the other map between the $(d+1)$-dimensional fields and the Carrollian fields
\be \label{elec-rescale-p}
F_{+i_1...i_{p}}[A] \longrightarrow  \tilde{F}_{+i_1...i_{p}} [A_e]\,,  \quad F_{i_1...i_{p+1}}[A]  \longrightarrow 0 \,,
\ee
we get the electric Carroll action for free $p$-form fields from the Bargmann action 
\be \label{S-Carr-elec-p}
 {}^{(d)}\mathcal S^{elec} = \lim_{\varepsilon \to 0} \mathcal S^{Barg}_{p\text{-form}} =- \frac{\alpha}{2(p+1)!} \int dx^+ d^\perp  x\,( \tilde{F}_{+i_1...i_{p}})^2  \,.
\ee
The procedure may also be extended to include interactions, as in the case of Yang-Mills theory discussed in the previous section.
\par
We illustrate this below with the examples of scalar fields ($p=0$) and electromagnetism ($p=1$), showing that this method yields the same Carrollian theories as those derived earlier using Hamiltonian techniques.
\par
For scalars $(p=0)$, the field strength is simply $F_\mu =  \del_\mu \phi$. Therefore, the Bargmann-invariant action \eqref{S-Barg-sc} can be recast into the form
\be
\mathcal S^{Barg}_{Scalar} = \int dx^+ dx^- d^\perp x \left( \frac{1}{2} {\alpha}\, \mathfrak n^\mu  \mathfrak n^\nu F_{\mu} F_\nu -  \frac{1}{2} \eta^{\mu \nu } F_\mu F_\nu  \right) \,.
\ee
While performing the null reduction, we can set $F_+[\phi]  \to 0$ and $F_i [\phi]  \to F_i [\phi_m]$, which reproduces the Carroll magnetic action \eqref{S-Carr-mag} in the Lagrangian form
\be \label{S-mag-Lag}
\mathcal S^{mag}_{Scalar} =  \frac{1}{2}\int dx^+  d^\perp x \  F_{i} [\phi_m] F^i  [\phi_m] = \frac{1}{2}\int dx^+  d^\perp x\ \del_i \phi_m \del^i \phi_m \,.
\ee
Or else, we could set $F_+ [\phi]  \to F_+ [\phi_m]$ and $F_i [\phi] \to 0$, to obtain the electric Carroll action \eqref{S-Carr-elec-sc} in the Lagrangian form
\be \label{S-elec-Lag}
\mathcal S^{elec}_{Scalar} =  \frac{1}{2}\alpha \int dx^+  d^\perp x \  F_{+} [\phi_m] F_+  [\phi_m] = \frac{1}{2} \alpha\int dx^+  d^\perp x\ (\del_+ \phi_m)^2 \,.
\ee
\par 
Similarly, for electromagnetism $(p=1)$, we consider the Bargmann action \eqref{S-Barg-EM}
\be
\mathcal S^{Barg}_{EM} = \int dx^+ dx^- d^\perp x \left( \frac{1}{2} \alpha\, \mathfrak n^\mu \mathfrak n^\nu\eta^{\rho \sigma}_{lc}  F_{\mu \rho }F_{\nu \sigma} - \frac{1}{4} F_{\mu \nu } F^{\mu \nu}\right) \,.
\ee
We can set $F_{ij}  \to \tilde{F}_{ij}$ and $F_{+i} \to 0$ to reproduce the magnetic action \eqref{S-Carr-mag-EM} in the Lagrangian form
\be
\mathcal S^{mag}_{EM} = -\frac{1}{4}\int dx^+  d^\perp x \  F_{ij}  F^{ij} = -\frac{1}{2}\int dx^+  d^\perp x \ (\vec{B}_{Carr})^2 \,.
\ee
For the electric sector, we set $F_{ij}  \to 0 $ and $F_{+i} \to \tilde F_{+i}$ which yields the electric Carroll action \eqref{S-Carr-elec-EM}
\be
\mathcal S^{elec}_{EM} = \frac{1}{2} \alpha\int dx^+  d^\perp x \  F_{+i}  F_{+i} = \frac{1}{2} \alpha\int dx^+  d^\perp x \ (\vec{E}_{Carr})^2 \,,
\ee
where $E^{Carr}_i = \tilde F_{+i}$ and $B^{Carr}_i = \epsilon_{ijk}F^{jk}/2$, as in the case of Yang-Mills theory.
\par
An interesting point to note here is that in light-cone gauge theories, although the vector (or $p$-form) fields in $(d+1)$ dimensions are not manifestly Lorentz covariant, they retain the underlying Carroll covariance in one lower dimension. In the Lorentzian case, there is no covariant notion of an electric or magnetic field in light-cone coordinates. For instance, we can formally group certain components of the field strength $-$ say, in four dimensions $-$ to resemble the `electric' and `magnetic' 3-vectors 
\be
\vec{E}^{Lor} \sim \{F_{+-}, F_{+i}\} \,, \quad \vec{B}^{Lor} \sim\{F_{-i}, F_{ij}\} \,.
\ee
However, these do not transform covariantly as 3-vectors under the Lorentz group. When we perform the null reduction, their Carrollian counterparts, which are now $(d-1)$-vectors (or 2-vectors in the four-dimensional case), namely
\be
E^{Carr}_i = F_{+i}\,, \quad B^{Carr}_i =\frac{1}{2} \epsilon_{ijk}F^{jk}
\ee
do become the electric and magnetic fields in the $d$-dimensional Carrollian theory.

\section{Concluding remarks}
We conclude with some final remarks on the Carrollian theories derived from Lorentzian light-cone actions. In this work, we presented two distinct methods to obtain Carrollian field theories via null reduction of Lorentzian theories in one higher dimension, using the tools of both Hamiltonian and Lagrangian formulations. Our procedure offers an alternative route in which the magnetic sector emerges naturally, shedding light on various interesting properties of these theories.
\par
By directly reducing the Lorentzian action to a null plane, one obtains only the magnetic Carroll sector, while extracting the electric Carroll sector requires deforming the Lorentzian action. This does not imply that the original Lorentzian theory lacks an electric sector, but rather that the magnetic sector is intrinsically embedded within the light-cone formulation. The fact that we can obtain the magnetic  Carrollian theory without deforming the Lorentzian action hints at a deeper connection between light-cone dynamics and Carroll physics.
\par 
A further compelling argument supporting this connection comes from examining the solutions to the equations of motion in these theories. Using the example of scalar fields, we compare the equations of motion derived from the parent Lorentzian action \eqref{LC-Scalar}
\be \label{LorEOM}
  (-2 \del_- \del_+ - \del_i \del^i)\, \phi (x^+, x^-, x^i) = 0
\ee
with those derived from the magnetic and electric Carroll actions \eqref{S-mag-Lag}-\eqref{S-elec-Lag} 
\bea
 \del_i \del^i \phi_m (x^+, x^i) &=& 0 \,,  \label{magEOM} \\
 \alpha\, \del_+^2 \phi_e (x^+, x^i) &=& 0 \label{elecEOM} \,.
\eea
From the structure of these equations, we observe that a magnetic Carroll solution $\phi_m$ satisfying \eqref{magEOM} also solves the Lorentzian equation of motion \eqref{LorEOM} in one higher dimension, since $\del_- \phi_m = 0$ by construction. However, this is not true for the electric case, because solutions $\phi_e$ satisfying the electric Carroll equation \eqref{elecEOM} do not necessarily satisfy \eqref{LorEOM}. This implies that the $d$-dimensional magnetic Carroll sector represents a \textit{consistent truncation} of the Lorentzian solution space in $(d+1)$ dimensions, much like a Kaluza-Klein reduction~\cite{Duff:1985jd}, whereas the electric solutions do not.
\par Let us consider an expansion of the Lorentzian field near the null plane $x^- = x^-_0$
\be
\phi(x^+, x^- ,x^i) = \phi_0 (x^+, x^i) + \underbrace{\varepsilon \phi^{(1)}(x^+, x^- ,x^i) + \varepsilon^2 \phi^{(2)}(x^+, x^- ,x^i)+ \ldots}_{\text{higher modes}} \,,
\ee
where $\phi$ satisfies the field equation \eqref{LorEOM}. We note that for the `zero mode' $\phi_0$, the Lorentzian field equation \eqref{LorEOM} reduces to \eqref{magEOM}, indicating that it indeed behaves like a magnetic Carroll field. When $\varepsilon$ is very small, \ie\ , close to a constant null hypersurface, we can consistently discard all the higher modes keeping only the zero mode. Therefore, \textit{the behaviour of a Lorentzian light-cone field theory near a null hypersurface can be effectively modeled using a magnetic Carroll theory in one lower dimension}. The electric Carroll sector appears at order $\varepsilon$, but it is intertwined with the higher-order modes. As a result, it is not possible to consistently truncate the Lorentzian solution space to isolate only the electric Carroll sector.
\par 
In the light-cone formulation, a widely used technique known as discrete light-cone quantization (DLCQ) involves quantizing a theory within a finite light-cone interval or a spatial box of length $\varepsilon$~\cite{Heinzl:2000ht, Lenz:1991sa, Burkardt:1995ct}. In this approach, one restricts the domain of one of the coordinates, say $x^-$, to a finite width $\varepsilon$, which leads to a discretization of momenta and the energy spectra. Depending on the physical problem at hand, different limits of $\varepsilon$ are taken. For instance, in problems involving spontaneous symmetry breaking or zero modes, one considers the infinite volume or continuum limit $\varepsilon \to \infty$~\cite{Yamawaki:1998cy,Heinzl:1991ri}. Conversely, taking $\varepsilon \to 0$, which corresponds to sending the longitudinal momenta $p_- \to \infty$, yields the infinite momentum frame $-$ a limit that finds applications in ultraviolet analyses of quantum field theories and in string quantization~\cite{Brodsky:1997de,Banks:1996vh}.
\par
In this paper, we also consider the limit $\varepsilon \to 0$, albeit in a slightly different context. In fact, our procedure for deriving Carrollian field theories involves two parameters: the deformation parameter $\alpha$ and the width of the light-cone interval $\varepsilon$. As discussed earlier, $\alpha$ deforms the standard light-cone coordinates to the Bargmann coordinates $x^\pm_\alpha$, resulting in a general class of (flat) Bargmann metrics $G_{\mu \nu}$
\be
\eta_{\mu \nu}^{lc} \longrightarrow G_{\mu \nu} =\begin{pmatrix}
0&-1 & 0 \\
-1&-\alpha& 0 \\
0 & 0 & \delta_{ij}
\end{pmatrix} \,.
\ee
Out of the two vectors $\mathfrak n$ and $\mathfrak m$, only $\mathfrak n$ remains a null vector with respect to $G$, i.e., $\mathfrak n^\mu \mathfrak n^\nu G_{\mu \nu} = 0$, while $\mathfrak m^\mu  \mathfrak m^\nu G_{\mu \nu} = \alpha$. The light-cone Minkowski metric is thus recovered as a special case of $G_{\mu \nu}$ with $\alpha = 0$. Starting with a general Bargmann-invariant theory parametrized by $\alpha$, we can recover the Lorentzian theory in the limit $\alpha \to 0$. The Carrollian limit, on the other hand, corresponds to taking $\varepsilon \to 0$, effectively projecting the higher-dimensional parent theory onto a null hypersurface. In the usual infinite momentum or Galilean limit, one first compactifies the $x^-$ direction and then takes the radius of compactification $\varepsilon$ to zero, as is typically done in Kaluza-Klein reductions. Here, instead, we perform a \textit{longitudinal contraction} along the $x^-$ direction by inserting a delta function distribution peaked at a fixed value of $x^-$ and then taking the width of the distribution  $\varepsilon$ to zero, thereby yielding the Carrollian limit. 
\par 
In this work, we have restricted our analysis to scalar and gauge theories in the light-cone formulation on flat backgrounds. It would be interesting to extend these results to fermions and supersymmetric theories~\cite{Bergshoeff:2023vfd}. For the spin-two case, our procedure can be directly applied to the free massless theory, namely the Pauli-Fierz action. However, it would be more instructive to consider the general case of Einstein gravity and recover the Pauli-Fierz theory as a special case. Extending the analysis to curved backgrounds would require working with gravity in the double-null formulation~\cite{dInverno:1980kaa, Reisenberger:2007ku,Alexandrov:2014rta} and deforming it to a curved Bargmann geometry. From such a theory, one could derive both electric and magnetic Carroll gravities.
\par
The double-null formulation of gravity is also crucial in the context of asymptotic symmetries, the infrared structure of gravity and flat space holography~\cite{Barnich:2024vks,Bekaert:2024tkv}. In this regard, the light-cone Minkowski case presented here, comprising two intersecting light fronts, serves as a toy model that mimics the asymptotic structure of spacetime where future and past null infinities meet at spatial infinity. Another promising direction would be to investigate whether the magnetic Carroll gravity derived directly from Lorentzian light-cone gravity via the null reduction approach matches existing results in the literature~\cite{Hartong:2015xda,Henneaux:2021yzg,Campoleoni:2022ebj}. We leave these investigations for future work.

\section*{Acknowledgments}
We thank the anonymous referee for useful comments and suggestions on the manuscript.
We thank Glenn Barnich, Marc Henneaux, Henning Samtleben, and Simone Speziale for many fruitful discussions. Thanks are due to the members of the String Theory Group at the Indian Institute of Technology Kanpur for stimulating interactions during a visit. We are grateful to the Galileo Galilei Institute for their hospitality during the workshop titled ``From Asymptotic Symmetries to Flat Holography: Theoretical Aspects and Observable Consequences,'' where part of this work was completed.
\appendix
%\section{Kinematical subalgebras within light-cone Poincar\'e}

\section{Non-Lorentzian spacetimes from light-cone Minkowski } \label{null-reduce}
In Section 2, we discussed the Bargmann subalgebras that unify the two sets of Carroll and Galilei subalgebras found within the light-cone Poincar'e algebra. In this appendix, we show from a geometric point of view how the underlying Carrollian and Galilean structures emerge from the flat Bargmann structure of the light-cone Minkowski spacetime using the null reduction method~\cite{Duval:2014uoa}.

\subsection{Bargmann to Carrollian spacetime} \label{Carroll-mani}
Here, we discuss how to obtain a Carrollian manifold as a null hypersurface embedded into a Bargmann manifold. Let us consider a $(d+1)$-dimensional Lorentzian manifold $M^{d+1}$ with the metric $G$
\be
G_{\mu \nu} = \eta^{lc}_{\mu \nu} =\begin{pmatrix}
0&-1 & 0 \\
-1&0& 0 \\
0 & 0 & \delta_{ij}
\end{pmatrix} \, \quad \mu, \nu=  +, - , i \,,
\ee
where  we assume that $x^\mu$ are the local coordinates on $M^{d+1}$. Both the vector $\mathfrak n = \partial_+$ and $\mathfrak m= \partial_-$ satisfy
\be
 \mathfrak  n^\mu  \mathfrak  n^\nu G_{\mu\nu} = 0= \mathfrak  m^\mu  \mathfrak  m^\nu  G_{\mu \nu} \,.
\ee
Therefore, the light-cone Minkowski spacetime corresponds to two distinct Bargmann structures, $(M^{d+1}, G_{\mu \nu}, \mathfrak n^\mu)$ and $(M^{d+1},G_{\mu \nu},  \mathfrak m^\mu)$. In the Minkowski spacetime, the light fronts at $x^\pm = \text{constant}$ have a flat Carroll structure, as we show below.
\vskip 0.2cm
We illustrate the technique for the case of a constant $x^-$ hypersurface. The light front $\Sigma$ at $x^- = x^-_0$, where $x^-_0$ is a constant, may be viewed as the embedding $\Phi : \Sigma \hookrightarrow M$ given by~\cite{Katsuno}
\be
x^+(y) = y^+\,, \quad x^-(y) = x^-_0\,, \quad x^i(y) = y^i\,,
\ee
where $y^a = (y^+, y^i)$ are the local coordinates on $\Sigma^d$. The associated one-form, often called the clock 1-form in non-Lorentzian geometries, is given by 
\be
\tau_\mu = G_{\mu \nu} \mathfrak n^\mu \quad \Rightarrow \quad \tau = -dx^-\,.
\ee
One finds that ker $\tau$ defines a foliation on $M$. Each leaf of this foliation, say $\Sigma: x^- =x^-_0$, is endowed with a Carroll structure.
This hypersurface $\Sigma$ has a degenerate metric 
\be
g_{ab} = \begin{pmatrix}
0 & 0 \\
0  & \delta_{ij}
\end{pmatrix} \,, 
\ee
and a vector field $\mathfrak n^a = \del_+$, which lies in the kernel of $g_{ab}$ such that $\mathfrak n^a g_{ab} = 0$. Thus, we see that the light front indeed has a flat Carrollian structure $(\Sigma^d, g_{ab}, \mathfrak n^a)$.  The isometries of the Carroll manifold  given by
\be
L_\xi \mathfrak n^a = 0 \,, \quad  L_\xi g_{ab} =0 \quad  \Rightarrow \quad  \xi = \xi^+ (y^i) \del_+ + (\omega^i_j y^j + a^i) \del_i\,,
\ee
are indeed the Carroll transformations. Additionally, if we restrict $\xi$ to be linear in the coordinates, we get the finite-dimensional Carroll group
\be \label{Carroll}
\xi^+ = b_i y^i+ a^+\,, \quad \xi^i = \omega^i_j y^j + a^i\,, 
\ee
where $b_i$ are the Carroll boosts, $\omega_{ij}$ the spatial rotations and $(a^+, a^i)$ the spacetime translations.
\par
In order to define an `inverse' of $g_{ab}$, we consider at every point on $\Sigma$,  a second null vector $\lambda^\mu$, not parallel to $ \mathfrak n^\mu$ such that
\be
G_{\mu \nu} \lambda^\mu \lambda^\nu = 0\,, \quad G_{\mu \nu} n^\mu \lambda^\mu = 1\,.
\ee
The vector $\lambda^\mu$ is called the rigging of $\Sigma$ and, in the present case, reads $\lambda = -\del_- = - \mathfrak m$. We can now define a 1-form on $\Sigma$ as 
\be
{\lambda_a = G_{\mu \nu} \lambda^\mu e^\nu_a\,, \quad \lambda_a\mathfrak n^a  = 1 \quad \Rightarrow \quad \lambda = dy^+}\,.
\ee
The degenerate metric $g_{ab}$ cannot be inverted, but using $\lambda_a$ we can find a symmetric tensor $g^{ab}$ with the properties
\be
g_{ab}\,g^{bc} = \delta_a{}^c - \lambda_a \mathfrak n^c \,, \quad \lambda_a g^{ab} = 0 \,,
\ee
which has the form
\be
 g^{ab} = \begin{pmatrix}
0 & 0 \\
0  & \delta^{ij}
\end{pmatrix} \,.
\ee
The induced volume form on $\Sigma$ simply reads
\be
d\omega_\Sigma =   dy^+ \wedge dy^1 \ldots  \wedge dy^{(d-1)}\,.
\ee

 \subsection{Bargmann to Galilean spacetime}\label{Galilei-mani}
 We now discuss how one can obtain a Galilean  or a Newton-Cartan manifold from the flat Bargmann structure of light-cone Minkowski by projecting onto a constant $x^+$ plane, or equivalently by performing a dimensional reduction along a null vector field~\cite{Julia:1994bs}.
 
To go from the Bargmann spacetime \( B^{d+1} \) to a Galilean manifold \( N^d \), we project out the \( x^+ \) direction. Let us assume that $x^a = (x^-, x^i)$ are the local coordinates on $N^d$. As in the previous section, the clock one-form is defined by
\be
\tau_\mu = G_{\mu \nu } \mathfrak n^\nu\,.
\ee
Pulling this back to $N^d$ gives
\be
\tau_a = e^\mu_a \tau_\mu \quad \Rightarrow \quad \tau = -dx^-\,.
\ee
The inverse spatial metric  $h$ on $N$ takes the form
\be
h^{ab}  = \begin{pmatrix}
0 & 0 \\
0  & \delta^{ij}
\end{pmatrix} \,, \quad \tau_a h^{ab} = 0 \,.
\ee
The flat Galilean structure is then given by $(N^d, h^{ab}, \tau_a)$. The isometries of the Galilean manifold are
\be
L_\xi \tau = 0 \,, \quad  L_\xi h = 0 \,,
\ee
which leads to the general form
\be
\xi = \xi^- \del_- + (\omega^i_j x^j + a^i(x^-)) \del_i \,.
\ee
If we restrict $\xi$ to be linear in the coordinates, we get the finite Galilei transformations
\be \label{Carroll}
\xi^- = a^-\,, \quad \xi^i = \omega^i_j x^j + v^i x^- + a^i\,, 
\ee
where $v_i$ are the Galilean boosts, $\omega_{ij}$ the spatial rotations and $(a^-, a^i)$ the spacetime translations.

\section{Canonical analysis of Lorentzian light-cone theories} \label{canonical}
In this appendix, we present a detailed discussion of the Hamiltonian analysis of Lorentzian light-cone field theories. Using scalar fields and Maxwell theory as examples, we focus on the primary second-class constraints that are characteristic of the light-cone formulation. This analysis allows us to highlight the similarities and differences between the Lorentzian actions and the deformed actions, both in the presence and absence of gauge symmetry.

\subsection{Scalar field theory} \label{scalar-canonical}
We start with the Lorentzian light-cone Lagrangian for massless scalar fields
\be 
\mathcal L^{Lor}= -\frac{1}{2} \eta_{\mu \nu}^{lc} \del^\mu \phi \del^\nu \phi ~=~ \del_- \phi \del_+ \phi - \frac{1}{2} \del_i \phi \del^i \phi \,.
\ee Treating $x^+$ as (Carrollian) time, we define the conjugate momenta as
\be
\pi = \frac{\delta \mathcal L^{Lor}}{\delta(\del_+ \phi)} = \del_- \phi \,.
\ee
The equal-time Poisson brackets are
\bea  
\{ \phi (x), \pi (y) \} \Big|_{x^+= y^+} &=& \delta(x^- - y^-) \delta^\perp (x-y) \,.
\eea
The conjugate momenta $\pi$ are constrained 
\be
\chi= \pi - \del_- \phi \,,
\ee
where $\chi$ is a second-class primary constraint
\be \label{chi-chi}
\{ \chi (x), \chi(y)\} = -2 \del^x_- \delta(x-y) \delta^\perp (x-y)\, = C_{\chi \chi}.
\ee
These primary constraints found in the light-cone formulation behave differently from the usual first-class primary constraints that appear in gauge theories, such as the Gauss constraint in electromagnetism. We will return to this point in the next section.
\par To the canonical Hamiltonian density
\be 
\mathcal H_C = \pi \del_+ \phi - \mathcal L^{Lor} = \frac{1}{2} \del_i \phi \del^i \phi \,,
\ee
 we add a term involving a Lagrange multiplier $\lambda$, resulting in the primary Hamiltonian
\be
\mathcal H_{p} = \mathcal H_C + \lambda \chi = \frac{1}{2} \del_i \phi \del^i \phi + \lambda ( \pi -\del_- \phi ) \,.
\ee
The consistency condition for the constraint $\chi$ 
\be
\{ \chi(x), H_p (y) \} = 0 \quad \Rightarrow \quad \del_- \lambda = \frac{1}{2} \del_i \del^i \phi \,,
\ee
determines the Lagrange multiplier upto a zero mode in $x^-$
\be
\lambda (x^+,x^-, x^i) =  \frac{1}{2} \int_0^{x^-} dy^- \del_i \del^i \phi (x^+, y^-, x^i) + \lambda_0 (x^+, x^i) \,.
\ee
\par It is important to note that there is a ``zero-mode'' part of the primary constraint, without any dependence on $x^-$, 
\be
\chi_0 = \lambda_0 \pi_0\,, \quad \del_- \lambda_0 =0\,, \quad  \del_- \pi_0= 0\,,
\ee
which appears to be first-class
\be
 \{ \chi_0 (x)\,, \chi_0(y) \} = 0 \,.
\ee
Hence, one must be cautious when strongly imposing the constraint by eliminating $\pi$ for $\del_-\phi$ and reducing to the Lagrangian subspace, since one loses the information about the zero modes in this process\footnote{These zero modes also play an important role in some interesting physical phenomena such as boundary degrees of freedom in residual gauge symmetries, Nambu-Goldstone modes associated with spontaneous symmetry breaking, to name a few~\cite{Majumdar:2024rxg, Yamawaki:1998cy,Heinzl:1991ri}.}. But in many cases such as massless free theories, the zero mode is actually constrained by the so-called `zero-mode constraint' and thus, does not contribute to any dynamics~\cite{Yamawaki:1998cy, Heinzl:1991ri} (See also the appendices in~\cite{Alexandrov:2014rta, Adami:2023wbe}).
\par
In this extended phase space, the Hamiltonian action takes the form
\be  
\mathcal S_{E} [\phi, \pi, \lambda ] = \int dx^- dx^+  d^\perp x  \Big( \pi \del_+ \phi - \frac{1}{2} \del_i \phi \del^i \phi - \lambda \chi \Big) \,.
\ee
The equation of motions obtained from the action are as follows
\bea
 &&  \pi -\del_- \phi = 0\,, \\
&& \del_+ \phi - \lambda = 0 \,, \\
&& -\del_+ \pi - \del_- \lambda + \del_i \del^i \phi = 0 \,.
\eea
The constraint $\chi_\phi$ is retrieved from the equation of motion for $\lambda$. The canonical structure is given by 
\be
{}^{(d+1)}\Omega = \int dx^- d^\perp x \ d_V \pi \wedge d_V \phi
 \ee
where $d_V$ denote exterior derivative in phase space.

\subsection{Maxwell theory} \label{Maxwell-canonical}
Here, we discuss the canonical formulation of electromagnetism in the light-cone frame, based primarily on the analyses presented in~\cite{Hanson:1976cn, Steinhardt:1979it}. Note that we are not gauge fixing the theory at this stage, but rather reformulating Maxwell theory in light-cone coordinates.
Starting with the light-cone Maxwell action
\be
\mathcal{S}^{Lor}_{EM} =  \int dx^+ dx^- d^\perp x\ \, \mathcal L_{EM}\,,
\ee
with the Lagrangian
\be
\mathcal L_{EM} = -\frac{1}{4} F_{\mu \nu }F^{\mu \nu }  =\frac{1}{2} (F_{+-})^2 + F_{+i} F_{-i} - \frac{1}{4} F_{ij}F^{ij} \,,
\ee
we derive the conjugate momenta
\bea
\pi^+ = 0\,, \quad \pi^- = F_{+-} \,, \quad \pi^i = F_{-i} \,.
\eea
We note that only $\pi^-$, involving a $\del_+$ derivative, is dynamical, while $\pi^+$ and $\pi^i$ are constrained. The theory, therefore, has one first-class primary constraint
\be
C = \pi^+\,,
\ee
and one second-class primary constraints
\be
C_i = \pi_i - F_{-i} \,.
\ee
The first-class constraint is typical of gauge theories and leads to the Gauss constraint in Maxwell theory. In contrast, the second-class constraint is specific to the null-front Hamiltonian analysis and indicates that the light-cone theory is first order in time.
\par 
The canonical Hamiltonian density reads
\be
\mathcal H_c = \pi^\mu \dot{A}_\mu - \mathcal L_{EM} = \frac{1}{2} (\pi^-)^2 + (\pi^- \del_- + \pi^k \del_k) A_+ + \frac{1}{4} (F_{ij})^2 \,,
\ee
and the Poisson brackets are given by
\be
\{A_\mu (x), \pi^\nu (y) \}\Big|_{x^+ = y^+} = \delta^\nu_\mu\, \delta(x^- - y^-) \delta^\perp (x-y)  \,.
\ee 
We now turn to the consistency conditions on the constraints. With the primary Hamiltonian
\be
H_p = H_c + \int dx^- d^\perp x \left( u C + u^k C_k\right)
\ee
where $u$ and $u^k$ are Lagrange multipliers, we find that the consistency condition for second-class constraints $C_k$ reads
\be
\dot{C}_k = \{ C_k (x), H_p (y)\} = \del_k \pi^- + \del^i F_{ik} - 2 \del_- u^k = 0\,.
\ee
This condition does not lead to any secondary constraints, but instead determines the Lagrange multipliers $u^k$. However, the consistency condition for the first-class constraint $C$
\be
\dot{C} = \{ C(x), H_p(y)\} = \del_- \pi^- + \del_i \pi^i 
\ee
leads to a secondary constraint 
\be
\mathcal G = \del_- \pi^- + \del_i \pi^i \,,
\ee
which is simply the Gauss law in light-cone coordinates. Furthermore, we find that $\mathcal G $ is first-class and conserved in time
\be
 \{ \mathcal G (x), H_p(y)\}  = 0 \,.
\ee

Therefore, the Dirac-Bergmann constraint algorithm~\cite{Dirac:1950pj, Bergmann:1949zz} stops here, as there are no further constraints in the theory. We thus arrive at the extended Hamiltonian density (neglecting some surface terms)
\be
\mathcal H_{ext} =   \frac{1}{2} (\pi^-)^2  + \frac{1}{4} (F_{ij})^2 + u\, C + u^k C_k + (v- A_+)\, \mathcal{G}\,.
\ee
From the equations of motion
\be
\{ A_\mu, H_{ext}\}  =0 \,, \quad \{ \pi^\mu, H_{ext}\}  =0 \,,
\ee
 we can determine the rest of the Lagrange multipliers as
 \be
u = \dot{A}_+ \,, \quad v =0\,.
 \ee
We refer readers to~\cite{Hanson:1976cn} for a more detailed treatment of the constraints. We finally arrive at the Hamiltonian action for light-cone electromagnetism.
 \be
\mathcal{S}_{ext}= \int dx^+ dx^- d^\perp x \left\{ \pi^- \del_+ A_- + \pi^i \del_+ A_i - \frac{1}{2} (\pi^-)^2  -\frac{1}{4} (F_{ij})^2 - u^k C_k + A_+ \mathcal{G} \right\}\,.
 \ee
 \par 
We can restrict this Lorentzian action to a constant null hypersurface by inserting the smearing function $\delta_\varepsilon$ into the action and setting
\be
A_- = 0 \,, \quad \pi^- = 0  \,.
\ee
Then, assuming the $(d+1)$-dimensional Lorentzian fields are mapped to $d$-dimensional Carrollian fields as follows
\be
A_i \to A_i^m \,, \quad A_+ \to A_+^m\,, \quad \pi^i \to p^i_m\,,
\ee
 we end up with the magnetic Carroll action for electromagnetism
\be 
 {}^{(d)}\mathcal S^{mag} [A_i, p^i] =\int dx^+ d^\perp x \left( p^i \del_+ A_i -\frac{1}{4} F_{ij}F^{ij}- A_{+} \mathcal G_{Carr}\right)\,.
\ee
But, just as in the case of scalars, it is not possible to derive the electric Carroll sector of the Maxwell theory from the Lorentzian action directly.

%%%%%%%%%%%%%%%%%%%%%%%%%%%%%%%%%%%%%%%%%%%%%%%%%%%%%%%%%%%%%%%%%%%%%%%%%%%%%%%%%%%%%%%%%%%%%%%%%%%%%%%%%%%%%%%%%%%%%%%%

\end{document}